\newcommand{\fmc}{\boldsymbol{\mathcal{F}}_{\text{mc}}}
\newcommand{\fhmc}{\hat{\boldsymbol{\mathcal{F}}}_{\text{mc}}}

\documentclass[10pt]{IEEEtran}  
\usepackage[ruled,linesnumbered]{algorithm2e}
\usepackage{anyfontsize}
\usepackage{booktabs}
\usepackage{multirow,hhline}
\usepackage{comment}
\usepackage{amsmath}
\usepackage{amssymb}
\usepackage{graphicx}
\usepackage{cite}
\usepackage{dirtytalk}
\usepackage{xcolor}
\usepackage{tikz}

\DeclareMathOperator*{\argmax}{arg\,max}

\usepackage{accents}
\newlength{\dhatheight}
\newcommand{\doublehat}[1]{%
    \settoheight{\dhatheight}{\ensuremath{\hat{#1}}}%
    \addtolength{\dhatheight}{-0.35ex}%
    \hat{\vphantom{\rule{1pt}{\dhatheight}}%
    \smash{\hat{#1}}}}
\makeatletter
\newlength \figwidth
\if@twocolumn
\else
\fi
\makeatother
\title{Multi-Source DOA Estimation through Pattern Recognition of the Modal Coherence of a Reverberant Soundfield}
\author{
Abdullah Fahim, Prasanga N. Samarasinghe, Thushara D. Abhayapala\\
Audio \& Acoustic Signal Processing Group, The Australian National University, Canberra, Australia
\thanks{This work is supported by the Australian National University Strategic Research Funds.}
}
\newcommand\copyrighttext{%
  \footnotesize \textcopyright 2019 IEEE. Personal use of this material is permitted. Permission from IEEE must be obtained for all other uses, in any current or future media, including reprinting/republishing this material for advertising or promotional purposes, creating new collective works, for resale or redistribution to servers or lists, or reuse of any copyrighted component of this work in other works. DOI: 10.1109/TASLP.2019.2960734
  }
\newcommand\copyrightnotice{%
\begin{tikzpicture}[remember picture,overlay]
\node[anchor=south,yshift=0pt] at (current page.south) {\fbox{\parbox{\dimexpr\textwidth-\fboxsep-\fboxrule\relax}{\copyrighttext}}};
\end{tikzpicture}%
}
\begin{document}
\bstctlcite{IEEEexample:BSTcontrol}
\maketitle
\copyrightnotice
\thispagestyle{empty}
\pagestyle{empty}
%
%%%%%%%%%%%%%%%%%%%%%%%%%%%%%%%%%%%%%%%%%%%%%%%%%%%%%%%%%%%%%%%%%%%%%%%%%%%%%%%%
%
\begin{abstract}
We propose a novel multi-source direction of arrival (DOA) estimation technique using a convolutional neural network algorithm which learns the modal coherence patterns of an incident soundfield through measured spherical harmonic coefficients. We train our model for individual time-frequency bins in the short-time Fourier transform spectrum by analyzing the unique snapshot of modal coherence for each desired direction. The proposed method is capable of estimating simultaneously active multiple sound sources on a $3$D space using a single-source training scheme. This single-source training scheme reduces the training time and resource requirements as well as allows the reuse of the same trained model for different multi-source combinations. The method is evaluated against various simulated and practical noisy and reverberant environments with varying acoustic criteria and found to outperform the baseline methods in terms of DOA estimation accuracy. Furthermore, the proposed algorithm allows independent training of azimuth and elevation during a full DOA estimation over $3$D space which significantly improves its training efficiency without affecting the overall estimation accuracy.
\end{abstract}
%
%%%%%%%%%%%%%%%%%%%%%%%%%%%%%%%%%%%%%%%%%%%%%%%%%%%%%%%%%%%%%%%%%%%%%%%%%%%%%%%%
\section{Introduction} \label{lab: introduction}
\IEEEPARstart{T}{he} task of a direction of arrival (DOA) estimator is to identify sound source directions from the output of a microphone array. DOA estimation is typically a prerequisite for several signal processing algorithms such as beamforming, power spectral density (PSD) estimation, and spatial audio coding, which in turn are integral parts of many practical applications such as multi-source separation \cite{hioka2013underdetermined,nugraha2016multichannel,gannot2017consolidated}, speech recognition \cite{busso2005smart}, robot audition \cite{nakadai2000active,yamamoto2007design},  bio-diversity monitoring \cite{blumstein2011acoustic,chu2009environmental}, audio surveillance and smart home applications \cite{chen2013smart,crocco2016audio}. In many practical environments, we face a scenario where multiple sound sources from different directions contribute to the acoustic scene. In such mixed recording scenarios, different sound sources can overlap on each other partially (e.g., teleconference) or fully (e.g., cocktail party problem, bioacoustic) over time. In this paper, we utilize the modal coherence of the spherical harmonic coefficients of a reverberant soundfield to train a convolutional neural network (CNN) for DOA estimation of multiple sound sources irrespective of their overlapping nature.\par
\subsection{Literature review}
DOA estimation is a decade-old problem with a number of algorithms developed over the years to accurately estimate sound source locations. However, while different algorithms have shown their usefulness under certain environments, they all have their own constraints and limitations and hence, DOA estimation remains an active problem in acoustic signal processing. A large number of DOA estimation techniques have been developed in the parametric domain \cite{kim1996two}. There are subspace-based methods like multiple signal classification (MUSIC) \cite{schmidt1986multiple} or the estimation of signal parameters via rotational invariance technique (ESPRIT) \cite{roy1989esprit} which utilizes the orthogonality between the signal and noise subspaces to estimate the source DOAs. MUSIC algorithm was originally developed for narrowband signals, however, it has been extensively used with wideband processing using a frequency smoothing technique \cite{khaykin2009coherent} or by decomposing the signal into multiple narrowband subspaces \cite{wang1985coherent}. It is common knowledge that the performance of the subspace-based methods are susceptible to strong reverberation and background noise \cite{dibiase2001robust}. Recently a variation of MUSIC was proposed in \cite{birnie2019sound} to improve its robustness in a reverberant room assuming the prior knowledge of room coupling coefficients.\par
There also exist beamforming-based methods for DOA estimation where the output power of a beamformer is scanned in all possible directions to find out when it reaches the maximum. A popular formulation of the beamformer-based technique is the steered response power (SRP) method which formulates the output power as a sum of cross-correlations between the received signals. Dibiase proposed an improvement to SRP in \cite{dibiase2000high} using the phase transform (PHAT) variant of the generalized cross-correlation (GCC) model \cite{knapp1976generalized}. The beamforming-based methods experience degradation in their performance for closely-spaced sources due to the limitation of the spatial resolution. Furthermore, both subspace and beamforming based techniques require to scan for all possible DOA angles during the run time which can be both time and resource intensive. Several modifications have been proposed to reduce the computational cost of SRP-PHAT by replacing the traditional grid search with region-based search \cite{marti2013steered,nunes2014steered,lima2014volumetric,do2007real}, however, this increases the probability of missing a desired source in reverberant conditions.\par
Another group of parametric approaches to DOA estimation uses the maximum likelihood (ML) optimization with the statistics of the observed data which usually requires accurate statistical modeling of the noise field \cite{ye1995maximum,chen2002maximum,stoica1990maximum}. In more recent works, DOA estimation, posed as a ML problem, was separately solved for reverberant environments \cite{mandel2007algorithm,schwartz2016multi} and with unknown noise power \cite{schwartz2017doa} using expectation-maximization technique. A large number of localization techniques are based on the assumption of non-overlapping source mixture in the short-time Fourier transform (STFT) domain, known as W-disjoint orthogonality (WDO) \cite{yilmaz2004blind}. Li \textit{et al.} adopted a Gaussian mixture model to estimate source locations using ML method on the basis of WDO \cite{li2017multiple}. The sparsity of speech signals in the STFT domain was exploited in \cite{zhang2010two,liu2000localization} to localize broadside sources by mapping phase difference histogram of the STFT coefficients. The works in \cite{rafaely2017speaker,nadiri2014localization} imply sparsity on both signals and reflections to isolate time-frequency (TF) bins that contain only direct path contributions from a single source and subsequently estimate source DOAs based on the selected TF bins. Recently, there has been an increase in efforts for intensity-based approaches where both sound pressure and particle velocity are measured and used together for DOA estimation \cite{levin2010angular,moore2015direction,moore2017direction,hafezi2017augmented}.\par
Lately, the application of spatial basis functions, especially the spherical harmonics, is gaining researchers' attention in solving a wide variety of acoustic problems including DOA estimation. Among the works we have referred so far in this paper, \cite{moore2017direction,hafezi2017augmented,khaykin2009coherent,birnie2019sound} were implemented in the spherical harmonic domain. Tervo \textit{et al.} proposed a technique for estimating DOA of the room reflections based on a ML optimization using a spherical microphone array \cite{tervo2015direction}. Kumar \textit{et al.} implemented MUSIC and beamforming-based techniques with the spherical harmonic decomposition of a soundfield for nearfield DOA estimation in a non-reverberant environment \cite{kumar2016near}. A free-field model of spherical harmonic decomposition was used in \cite{hafezi20163d} to perform an optimized grid search for acoustic source localization. The spherical harmonics are the natural basis functions for spatial signal processing and consequently offers convenient ways for recognizing the spatial pattern of a soundfield. Furthermore, the spherical harmonic coefficients are independent of the array structure, hence, the same DOA algorithm can be used with different shapes and designs of sensor arrays as long as they meet a few basic criteria of harmonic decomposition \cite{abhayapala2002theory,li2007flexible,chen2015theory,meyer2002highly,samarasinghe2017planar,abhayapala2010spherical}.\par
Over the past decade, the rapid technology advances in storage and processing capabilities led researchers to lean towards machine learning in solving many practical problems including DOA estimation. Being a data-driven approach, neural networks can be trained for different acoustic environments and source distributions. In the area of single source localization, significant progresses have been made in solving the limitations in the parametric approaches by incorporating machine learning-based algorithms. The authors of \cite{xiao2015learning,vesperini2016neural,sun2017indoor} derived features from different variations of the GCC model to train a neural network for single source localization. Ferguson \textit{et al.} used both cepstrogram and GCC to propose a single source DOA estimation technique for under-water acoustics \cite{ferguson2018sound}. Inspired by the MUSIC algorithm, the authors of \cite{takeda2016sound} utilized the eigenvectors of a spatial correlation matrix to train a deep neural network. Conversely, multi-source localization poses a more challenging problem to solve, especially with overlapping sources. In the recent past, a few algorithms have been proposed for multi-source localization based on CNN. A CNN-based multi-source DOA estimation technique was proposed in \cite{chakrabarty2019multi} where the authors used the phase spectrum of the microphone array output as the learning feature. The method in \cite{chakrabarty2019multi} was implemented in the STFT domain and all the STFT bins for each time frame were stacked together to form the feature snapshot. On the contrary, Adavanne \textit{et al.} considered both magnitude and phase information of the STFT coefficients and used consecutive time frames to form the feature snapshot to train a convolutional recurrent neural network (CRNN) and performed a joint sound event detection and localization \cite{adavanne2018sound}.  Both \cite{chakrabarty2019multi} and \cite{adavanne2018sound} require the model to be trained for  unique combinations of sound sources from different angles in order to accurately estimate the DOA of simultaneously active multiple sound sources.
\subsection{Contribution of this paper}
In this paper, we propose a CNN-based framework to estimate DOAs of simultaneously active multiple sound sources. We use a novel feature to train a CNN which utilizes the modal coherence of a reverberant soundfield in the spherical harmonic domain. We show that modal coherence represents a unique pattern for each direction which can be learned and used for estimating source DOAs in a composite acoustic scene. Note that, previously we developed a parametric multi-source separation algorithm in \cite{fahim2018psd} using the modal coherence model, hence, the proposed method offers an efficient way of joint multi-source localization and separation in a reverberant environment.\par
We design the algorithm to perform multi-source DOA estimation while being trained for the single-source case only. To the best of our knowledge, this approach has not been taken in the learning-based DOA estimation so far as most of the existing techniques require the training data to contain different angular combinations from the desired DOA grid. The single-source training model of the proposed method offers multiple advantages during training and testing stages. As the number of unique combinations increases rapidly with the number of audio sources and size of the DOA grid\footnote{For a perspective, $250$ distinct DOA points contribute to approximately $2.5$ million and $160$ million unique combinations for $3$ and $4$ source-mixtures, respectively.}, a requirement for multi-source training causes a significant increment of time and resource requirements compared to the single-source training model. Furthermore, the adoption of single-source training allows us to train the model for once and use the same model in the testing environment irrespective of the number of sound sources in the mixed acoustic scenario.\par
Unlike the existing methods in the CNN domain, we treat individual STFT bins separately for training. The proposed algorithm uses the W-disjoint orthogonality assumption \cite{yilmaz2004blind} for multi-source separation and proposes a probability-based post-filtering of the CNN output to address any violation of W-disjoint orthogonality. Furthermore, as the training stage of the proposed algorithm involves single-source scenarios only, we can independently train our model for azimuths and elevations based on the same input dataset provided that we measure the soundfield for various source positions in each intended azimuth and elevation planes. Hence, we can share the same convolutional layers and perform joint azimuth and elevation estimation with two separate fully connected heads. In the results section, we include a demonstration for such a joint estimation along with a complete performance evaluation of the proposed method with a wide range of criteria such as different practical and simulated room conditions with a variable number of sources. We also compare and analyze the results with another state of the art CNN-based algorithm for DOA estimation.\par
The remainder of the paper is structured as follows. Section \ref{sec: problem-formulation} contains the problem statement and defines the objective of the work. In Section \ref{sec: modal-coherence}, we briefly describe the existing frameworks for spherical harmonic decomposition and modal coherence of a reverberant soundfield. Section \ref{sec: cnn-doa-estimation} contains a detailed description of the proposed model including different aspects of feature selection. Finally, in Section \ref{sec: experimental-results}, we evaluate and analyze the performance of the proposed algorithm and compare it with a contemporary method based on objective metrics and graphical aid.
\begin{figure}[!t]
\centering
\begin{minipage}[b]{0.9\linewidth}
  \centering
  \centerline{\includegraphics[width=\linewidth]{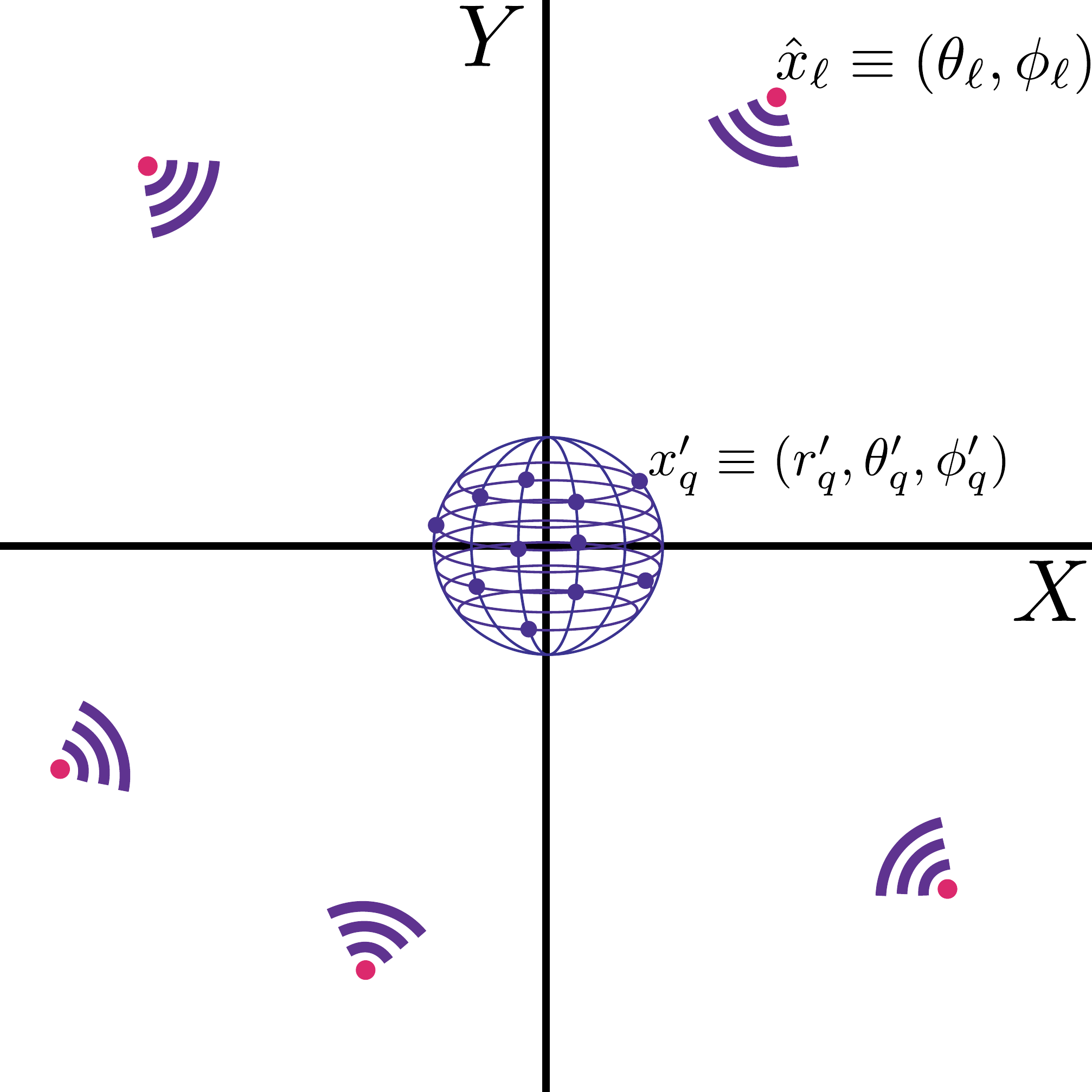}}
\end{minipage}
\caption{A graphical impression of a spherical microphone array setup in the presence of multiple sound sources. Array shape may differ depending on the spherical harmonic decomposition technique.}
\label{fig: source-mic-arrangement}
\end{figure}
\section{Problem formulation} \label{sec: problem-formulation}
Consider $L$ sound sources concurrently emitting sound in a reverberant room. The sound pressure observed by an omnidirectional microphone placed at a coordinate $\boldsymbol{x}^{\prime}_q \equiv (r^{\prime}_q, \theta^{\prime}_q, \phi^{\prime}_q)$ inside the room, where $r^{\prime}_q$, $\theta^{\prime}_q$, and $\phi^{\prime}_q$ are the radius, elevation, and azimuth of point $\boldsymbol{x}^{\prime}_q$ in the spherical coordinate system, respectively, is expressed by
\begin{equation} \label{eq: mixed-sound-td}
p(\boldsymbol{x}^{\prime}_q, t) = \sum \limits_{\ell=1}^{L} h_{\ell}(\boldsymbol{x}^{\prime}_q, t) \ast s_{\ell}(t)
\end{equation}
where $t$ is the discrete time index, $h_{\ell}(\boldsymbol{x}^{\prime}_q, t)$ is the room impulse response (RIR) between the $\ell^{th}$ source position and $\boldsymbol{x}^{\prime}_q$, $s_{\ell}(t)$ is the $\ell^{th}$ source signal, and $\ast$ denotes the convolution operation. The corresponding frequency domain representation of \eqref{eq: mixed-sound-td} in STFT domain can be obtained using the multiplicative model of convolution and is formulated as
\begin{equation} \label{eq: mixed-sound-fd}
    P(\boldsymbol{x}^{\prime}_q, k, \tau) = \sum \limits_{\ell = 1}^{L} S_{\ell}(k, \tau) H_{\ell}(\boldsymbol{x}^{\prime}_q, k)
\end{equation}
where $\{P, S, H\}$ represent the corresponding signals of $\{p, s, h\}$ in the STFT domain, $\tau$ is the timeframe index, $k = 2 \pi f / c$, $f$ denotes the frequency, and $c$ is the speed of sound propagation. Henceforth, $\tau$ is omitted for brevity as we shall treat each of the time frames independently.\par
In this work, we intend to estimate the individual DOAs in the presence of multiple concurrent sound sources, i.e., we want to estimate $\doublehat{\boldsymbol{x}}_{\ell} \equiv (\hat{\theta}_{\ell}, \hat{\phi}_{\ell}) \text{ } \forall \ell \in [1, L]$, given a set of measured sound pressure $p(\boldsymbol{x}^{\prime}_q, t) \text{ } \forall q \in [1, Q]$ or the corresponding spherical harmonic coefficients\footnote{The spherical harmonic decomposition technique is described in Section \ref{sec: spherical-harmonics-decomposition}.} of a mixed soundfield. We pose the DOA estimation as a CNN classification problem where we sample the intended DOA range into discrete sets $\Theta = \big \{\theta_a \big \}_{a \in [1, I]}$ for elevations and $\Phi = \big \{\phi_b \big \}_{b \in [1, J]}$ for azimuths. Thereafter, we propose a feature unique to each angle and train a CNN framework individually for each of the members of $\Theta$ and $\Phi$. Finally, during the evaluation, the CNN model finds the closest match of the true DOA $\hat{\boldsymbol{x}}_{\ell} \equiv (\theta_{\ell}, \phi_{\ell})$ $\forall \ell$ in the DOA sets $\Theta$ and $\Phi$ based on its learning and accurately combines the independent estimations $\hat{\theta}_{\ell}$ and $\hat{\phi}_{\ell}$ for each individual source to achieve full DOA estimation.
\section{Modal framework} \label{sec: modal-coherence}
In this section, we describe a few established concepts in the spherical harmonic domain that is used as the framework for the proposed DOA estimation technique.
\subsection{Spherical harmonic decomposition of a soundfield} \label{sec: spherical-harmonics-decomposition}
A continuous soundfield on a sphere can be decomposed using the spherical harmonic basis functions as \cite{williams1999fourier}
\begin{equation} \label{eq: sh-decomposition}
    P(\boldsymbol{x}^{\prime}_q, k) = \sum \limits_{nm}^{\infty} \alpha_{nm}(k) \text{ } b_n(k r^{\prime}_q) \text{ } Y_{nm}(\hat{\boldsymbol{x}}^{\prime}_q)
\end{equation}
where $\sum \limits_{nm}^{(\cdot)} \equiv \sum \limits_{n=0}^{(\cdot)} \sum \limits_{m=-n}^{n}$, $\alpha_{nm}(k)$ is the sensor-independent soundfield coefficient of order $n$ and degree $m$, the position vector $\boldsymbol{x}^{\prime}_q \equiv (r^{\prime}_q, \hat{\boldsymbol{x}}^{\prime}_q)$, and the unit vector $\hat{\boldsymbol{x}}^{\prime}_q \equiv (\theta^{\prime}_q, \phi^{\prime}_q)$. The infinite summation of \eqref{eq: sh-decomposition} is often truncated at the soundfield order $N = \lceil kr^{\prime}_q \rceil$ \cite{jones2002dimensionality,ward2001reproduction}, where $\lceil \cdot \rceil$ denotes the ceiling operation, due to the high-pass nature of the higher-order Bessel functions. The complex spherical harmonic basis function $Y_{nm}(\cdot)$ is defined as
\begin{equation} \label{eq: sh-definition}
    Y_{nm}(\hat{\boldsymbol{x}}^{\prime}_q) = \sqrt{\frac{(2n +1)}{4 \pi} \frac{(n-\lvert m \rvert)!}{(n+\lvert m \rvert)!}} \text{ } \mathcal{P}_{n \lvert m \rvert} \big( \cos \theta^{\prime}_q \big) \text{ } e^{i m \phi^{\prime}_q}
\end{equation}
where $\lvert \cdot \rvert$ denotes absolute value, $(\cdot)!$ represents factorial, $\mathcal{P}_{n \lvert m \rvert}(\cdot)$ is an associated Legendre polynomial, and $i = \sqrt[]{-1}$. Furthermore, the dependency on array radius comes through the function $b_n(\cdot)$ which is defined as
\begin{equation} \label{eq: bn-definition}
b_n(\xi) = 
\begin{cases}
j_n(\xi) & \text{for an open array} \\
j_n(\xi) - \frac{j'_n(\xi)}{h'_n(\xi)} h_n(\xi) & \text{for a rigid spherical array}.
\end{cases}
\end{equation}
The spherical harmonic coefficients $\alpha_{nm}$ can be estimated using different kinds of arrays where the process and the formulation depend on the array geometry. E.g., utilizing the orthonormal property of the spherical harmonics, a spherical microphone array allows us to calculate $\alpha_{nm}$ from \eqref{eq: sh-decomposition} as \cite{abhayapala2002theory,meyer2002highly}
\begin{align}
    \alpha_{nm}(k) &= \frac{1}{b_n(kr)} \text{ } \int_{\mathbb{S}^2} \text{ } P(\boldsymbol{x}, k) \text{ } Y_{nm}^{*}(\hat{\boldsymbol{x}}) \text{ } d\hat{\boldsymbol{x}}  \label{eq: alpha-theroetical} \\
    & \approx \frac{1}{b_n(kr)} \sum \limits_{q=1}^{Q} w_q \text{ } P(\boldsymbol{x}^{\prime}_q, k) \text{ } Y_{nm}^{*}(\hat{\boldsymbol{x}}^{\prime}_q) \label{eq: alpha-practical}
\end{align}
where $r$ is the array radius, $\boldsymbol{x} = (r, \hat{\boldsymbol{x}})$, and $\hat{\boldsymbol{x}}$ denotes an arbitrary direction on the spherical shell $\mathbb{S}^2$. Obviously, it is impractical to realize a spatially continuous microphone array as required by \eqref{eq: alpha-theroetical}, hence, \eqref{eq: alpha-practical} is used as an approximation of \eqref{eq: alpha-theroetical} with $Q$ microphones. $w_q \text{ } \forall q$ are suitable microphone weights that ensure the validity of the orthonormal property of the spherical harmonics with a limited number of sampling points, i.e.,
\begin{equation}\label{eq: sh-orthonormality-band-limited}
\sum \limits_{q=1}^{Q} w_q \text{ } Y_{nm}(\boldsymbol{\hat{x}}^{\prime}_q) \text{ } Y^*_{n'm'}(\boldsymbol{\hat{x}}^{\prime}_q) \approx \delta_{nn'} \delta_{mm'}
\end{equation}
where $\delta_{nn'}$ and $\delta_{mm'}$ are the Kronecker delta functions.\par
The same spherical harmonic decomposition can be achieved using alternate array geometries and formulation, \cite{abhayapala2002theory,li2007flexible,chen2015theory,meyer2002highly,samarasinghe2017planar,abhayapala2010spherical} are a few examples of such procedures.
\subsection{RTF in the spatial domain}
The room transfer function (RTF) can be decomposed into two parts
\begin{equation} \label{eq: rir-parts}
    H_{\ell}(\boldsymbol{x}^{\prime}_q, k) = H_{\ell}^{(d)}(\boldsymbol{x}^{\prime}_q, k) + H_{\ell}^{(r)}(\boldsymbol{x}^{\prime}_q, k)
\end{equation}
where $H_{\ell}^{(d)}(\cdot)$ and $H_{\ell}^{(r)}(\cdot)$ are the corresponding direct and reverberant path components of the RTF. The RTF components are modeled in the spatial domain as
\begin{equation} \label{eq: rtf-modal-direct}
H^{(d)}_{\ell}(\boldsymbol{x}^{\prime}_q, k) = G^{(d)}_{\ell}(k) \text{ } e^{ik \text{ } \hat{\boldsymbol{x}}_{\ell} \cdot \boldsymbol{x}^{\prime}_q}
\end{equation}
\begin{equation} \label{eq: rtf-modal-reverb}
H^{(r)}_{\ell}(\boldsymbol{x}^{\prime}_q, k) = \int_{\mathbb{S}^2} G^{(r)}_{\ell}(k, \hat{\boldsymbol{x}}) \text{ } e^{ik \text{ } \hat{\boldsymbol{x}} \cdot \boldsymbol{x}^{\prime}_q} \text{ } d\hat{\boldsymbol{x}}
\end{equation}
where $G^{(d)}_{\ell}(k)$ represents the direct path gain between the origin and the $\ell^{th}$ source and $G^{(r)}_{\ell}(k, \hat{\boldsymbol{x}})$ is the reflection gain at the origin along the direction of $\hat{\boldsymbol{x}}$ for the $\ell^{th}$ source. Hence, we obtain the spatial domain equivalent of \eqref{eq: mixed-sound-fd} by substituting the spatial domain RTF from \eqref{eq: rir-parts} - \eqref{eq: rtf-modal-reverb} as
\begin{multline} \label{eq: mixed-sound-sd}
P(\boldsymbol{x}^{\prime}_q, k) = \sum \limits_{\ell = 1}^{L} S_{\ell}(k) \Bigg( G^{(d)}_{\ell}(k) \text{ } e^{ik \text{ } \hat{\boldsymbol{x}}_{\ell} \cdot \boldsymbol{x}^{\prime}_q} \\
+ \int_{\mathbb{S}^2} G^{(r)}_{\ell}(k, \hat{\boldsymbol{x}}) \text{ } e^{ik \text{ } \hat{\boldsymbol{x}} \cdot \boldsymbol{x}^{\prime}_q} \text{ } d\hat{\boldsymbol{x}} \Bigg).
\end{multline}
\subsection{Modal coherence model}
The spherical harmonic expansion of the far-field approximation of the Green's function is given by \cite[pp. 27--33]{colton2012inverse}
\begin{equation} \label{eq: far-field-source-green-func}
e^{ik \text{ } \boldsymbol{\hat{x}}_{\ell} \cdot \boldsymbol{x}^{\prime}_q} = \sum \limits_{nm}^{\infty} \text{ } 4 \pi i^{n} \text{ } Y^*_{nm}(\boldsymbol{\hat{x}}_{\ell}) \text{ } b_n(kr) \text{ } Y_{nm}(\boldsymbol{\hat{x}}^{\prime}_q).
\end{equation}
Using \eqref{eq: far-field-source-green-func} in \eqref{eq: mixed-sound-sd} and then by comparing it with \eqref{eq: sh-decomposition}, we obtain an analytical expression for $\alpha_{nm}$ in a reverberant room as \cite{fahim2018psd}
\begin{multline} \label{eq: alpha-reverb}
\alpha_{nm} (k) = 4 \pi i^{n} \text{ } \sum \limits_{\ell=1}^{L} S_{\ell}(k)  \Bigg( G^{(d)}_{\ell}(k) \text{ } Y^*_{nm}(\hat{\boldsymbol{x}}_{\ell}) \\
+ \int_{\mathbb{S}^2} G^{(r)}_{\ell}(k, \hat{\boldsymbol{x}}) \text{ } Y^*_{nm}(\hat{\boldsymbol{x}}) \text{ } d\hat{\boldsymbol{x}} \Bigg).
\end{multline}
Taking into account the autonomous behavior of the reflective surfaces in a room (i.e., the reflection gains from the reflective surfaces are independent) and imposing the assumptions of uncorrelated sources, we previously established \cite{fahim2018psd} a closed form expression for the modal coherence of a reverberant soundfield as 
\begin{multline} \label{eq: coherence-model}
\mathbb{E} \Big\{ \alpha_{nm}(k) \alpha^*_{n'm'}(k) \Big\} = \sum \limits_{\ell=1}^{L} \mathbb{E} \bigg\{ \Big\lvert S_{\ell}(k) \Big \rvert ^2 \bigg\} \Bigg( \Upsilon_{nm}^{n'm'}(\hat{\boldsymbol{x}}_{\ell}) \\
\mathbb{E} \bigg\{ \Big \lvert G_{\ell}^{(d)}(k) \Big \rvert ^2 \bigg\} + 
\sum \limits_{vu}^{V} \mathbb{E} \Big\{ \gamma^{(\ell)}_{vu}(k) \Big\} \text{ } \Psi_{n,n',v}^{m,m',u} \Bigg)
\end{multline}
where $\mathbb{E}\{\cdot\}$ denotes expected value and
\begin{equation} \label{eq: reverb-coefficients}
\sum \limits_{vu}^{V} \text{ } \mathbb{E} \Big\{ \gamma^{(\ell)}_{vu}(k) \Big\} \text{ } Y_{vu}(\hat{\boldsymbol{x}}) = \mathbb{E} \bigg\{ \Big \lvert G^{(r)}_{\ell}(k, \hat{\boldsymbol{x}}) \Big \rvert ^2 \bigg\}
\end{equation}
\begin{equation} \label{eq: direct-path-upsilon}
\Upsilon_{nm}^{n'm'}(\hat{\boldsymbol{x}}_{\ell}) = C_{nn'} \text{ } Y^*_{nm}(\hat{\boldsymbol{x}}_{\ell}) \text{ } Y_{n'm'}(\hat{\boldsymbol{x}}_{\ell})
\end{equation}
\begin{equation} \label{eq: Psi}
\Psi_{n,n',v}^{m,m',u} = C_{nn'} \text{ } W_{v, n, n'}^{u, m, m'}
\end{equation}
\begin{equation} \label{eq: Cnn}
    C_{nn'} = 16 \pi^2 i^{n-n'}
\end{equation}
\begin{multline} \label{eq: Wvnn}
W_{v, n, n'}^{u, m, m'} = (-1)^m \text{ } \sqrt[]{\frac{(2v+1)(2n+1)(2n'+1)}{4 \pi}} \times \\
\left(\begin{array}{clcr}
v & n & n'\\
0 & 0 & 0  \end{array}\right) \text{ }
\left(\begin{array}{clcr}
v & n & n'\\
u & -m & m'  \end{array}\right)
\end{multline}
where $(\cdot)$ in \eqref{eq: Wvnn} represents Wigner-3j symbol \cite{olver2010nist}. Note that, though \eqref{eq: coherence-model} was developed using a far-field sound propagation model, it can easily be written for near-field sound sources by replacing \eqref{eq: far-field-source-green-func} with its near-field counterpart \cite{fahim2018psd}. Furthermore, for the temporal processing, it is common to estimate the expected value by applying the exponential moving average technique on the instantaneous measurements, i.e.,
\begin{multline} \label{eq: expected-value-mc}
\mathbb{E} \Big\{ \alpha_{nm}(k, \tau) \alpha^*_{n'm'}(k, \tau) \Big\} = (1 - \beta) \text{ } \alpha_{nm}(k, \tau) \times \\
\alpha^*_{n'm'}(k, \tau)
+ \beta \text{ } \mathbb{E} \Big\{ \alpha_{nm}(k, \tau-1) \text{ } \alpha^*_{n'm'}(k, \tau-1) \Big\} 
\end{multline}
where $\beta \in [0, 1]$ is a smoothing factor.
\section{CNN-based DOA estimation} \label{sec: cnn-doa-estimation}
CNN is a popular technique in the deep learning domain, and is predominantly used in computer vision applications. The input, often a $2$D or $3$D tensor, goes through multiple convolution filters followed by a traditional fully-connected neural network. In this work, we pose the DOA estimation problem as an image-classification problem where the input image represents the modal coherence of the soundfield.
\begin{figure*}[!ht]
\begin{minipage}[b]{0.24\linewidth}
  \centering
  \centerline{\includegraphics[width=\linewidth]{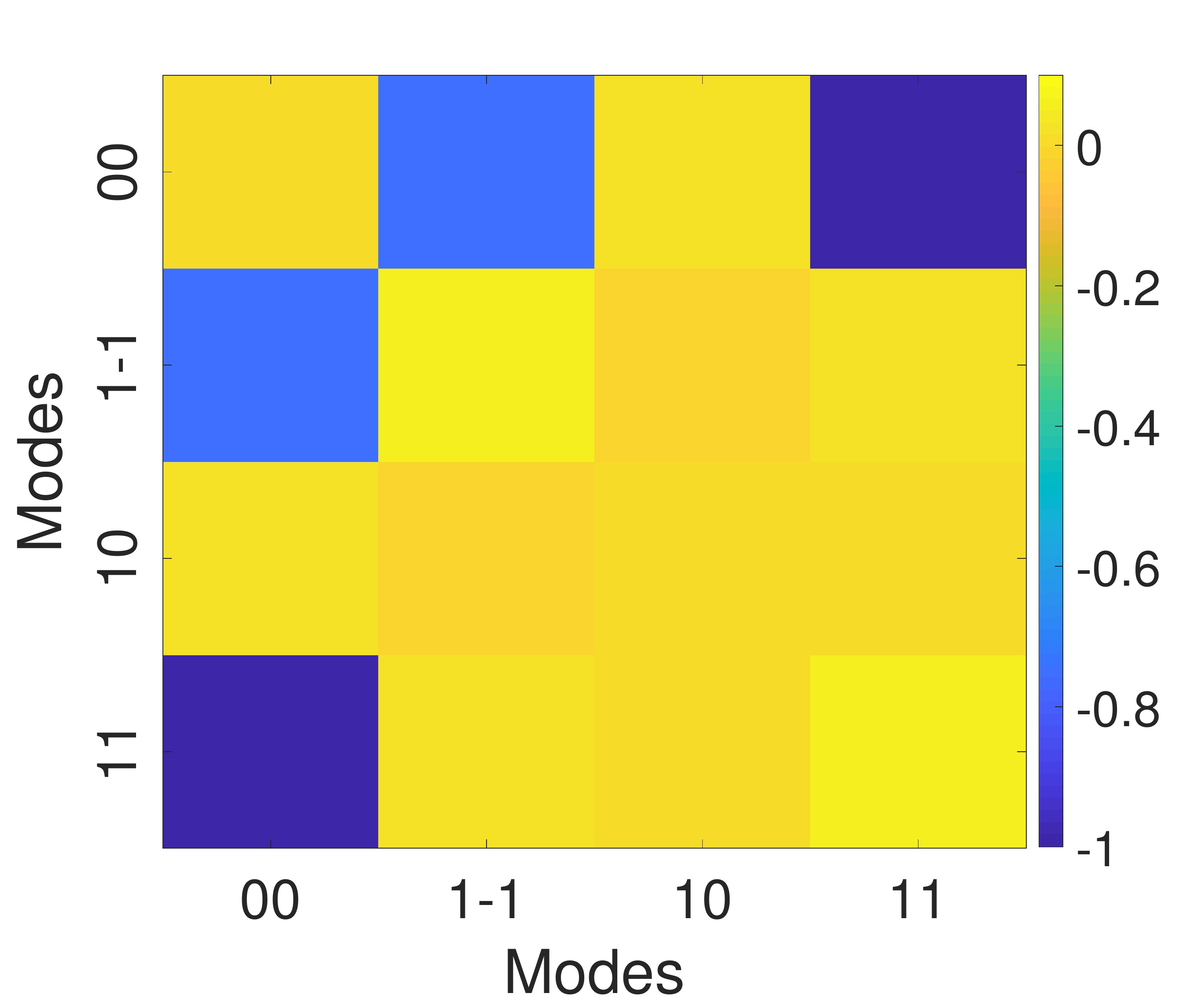}}
  \centerline{{\fontsize{9}{10}\selectfont(a)}}\medskip
\end{minipage}
\begin{minipage}[b]{0.24\linewidth}
  \centering
  \centerline{\includegraphics[width=\linewidth]{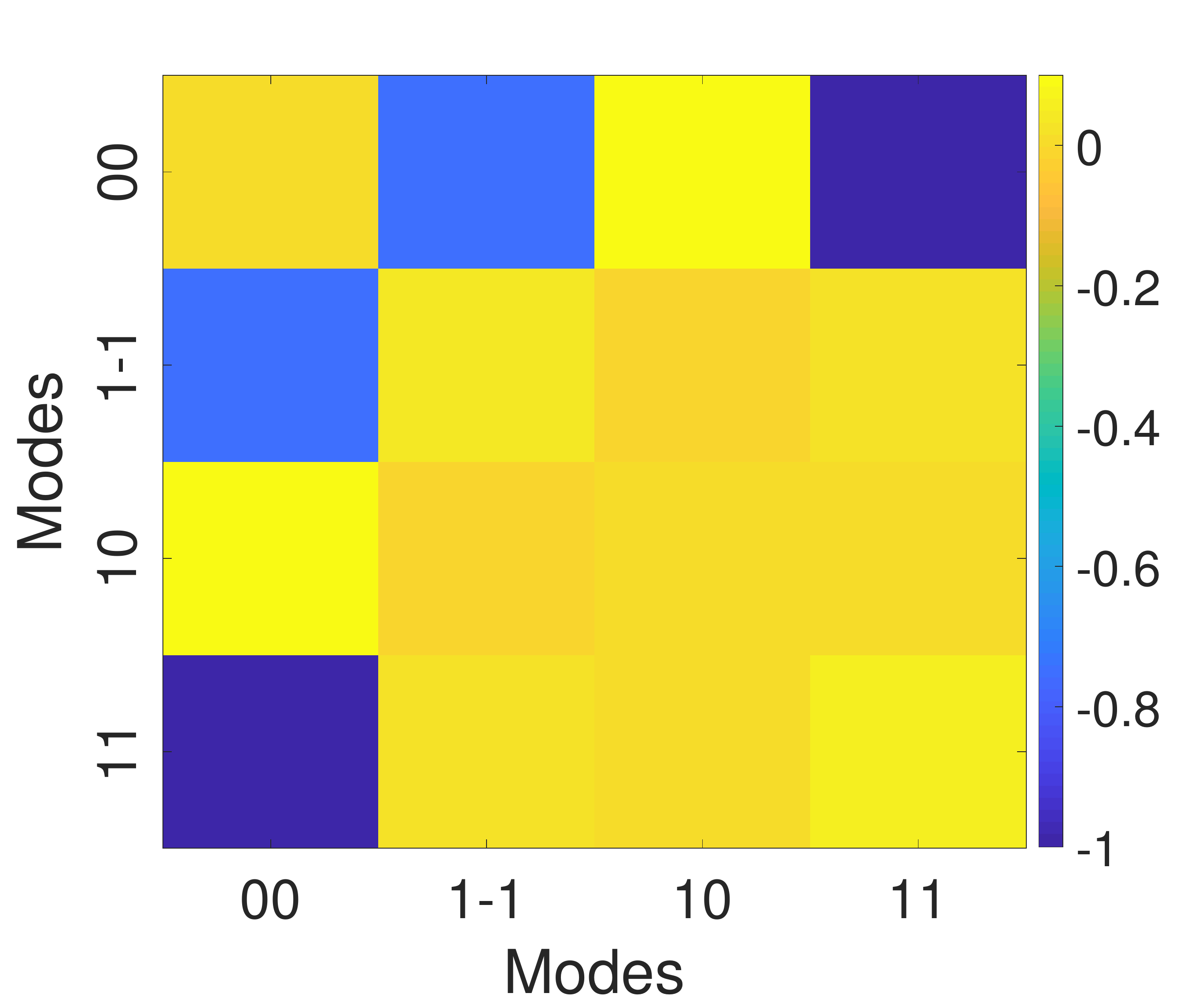}}
  \centerline{{\fontsize{9}{10}\selectfont(b)}}\medskip
\end{minipage}
\begin{minipage}[b]{0.24\linewidth}
  \centering
  \centerline{\includegraphics[width=\linewidth]{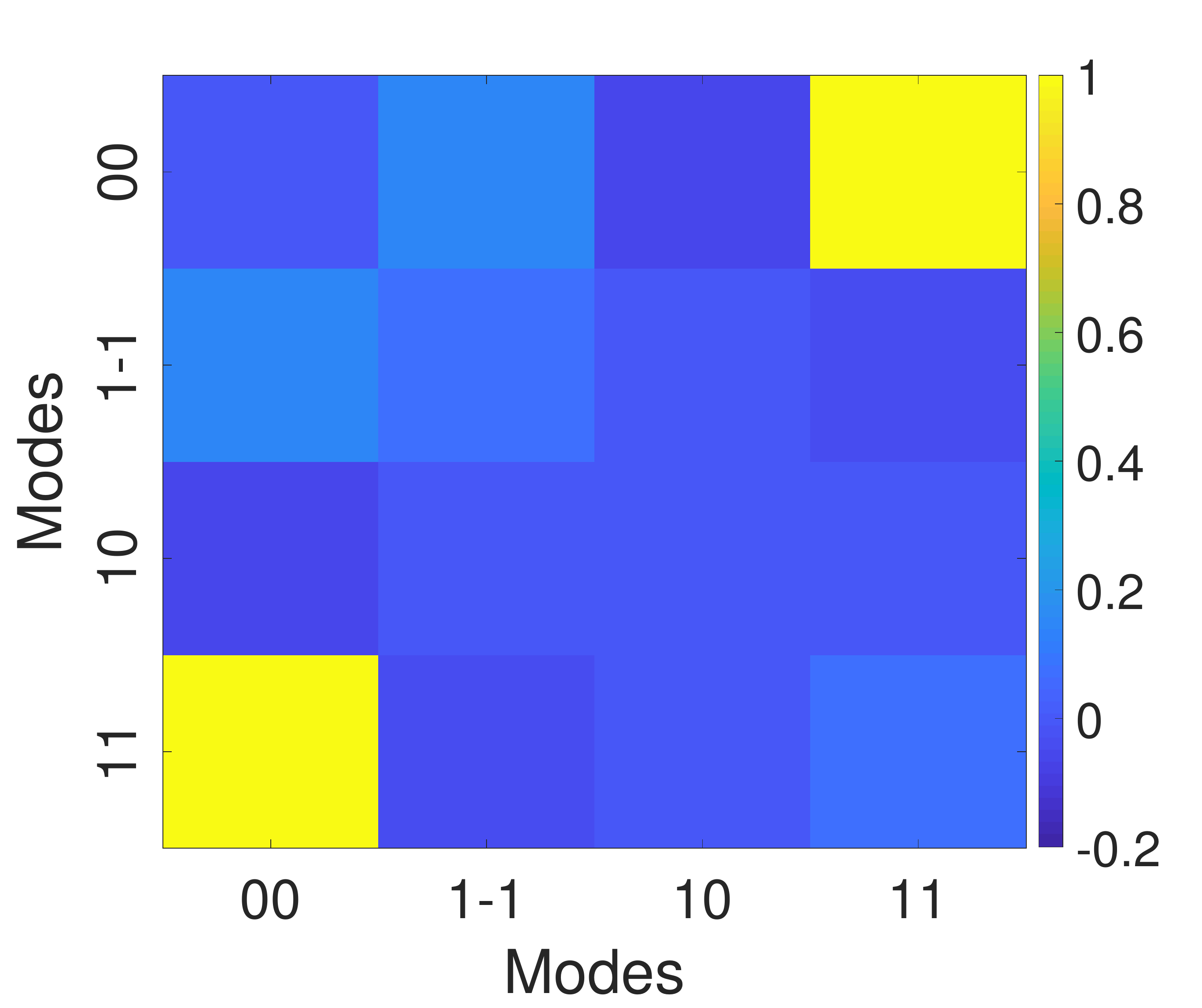}}
  \centerline{{\fontsize{9}{10}\selectfont(c)}}\medskip
\end{minipage}
\begin{minipage}[b]{0.24\linewidth}
  \centering
  \centerline{\includegraphics[width=\linewidth]{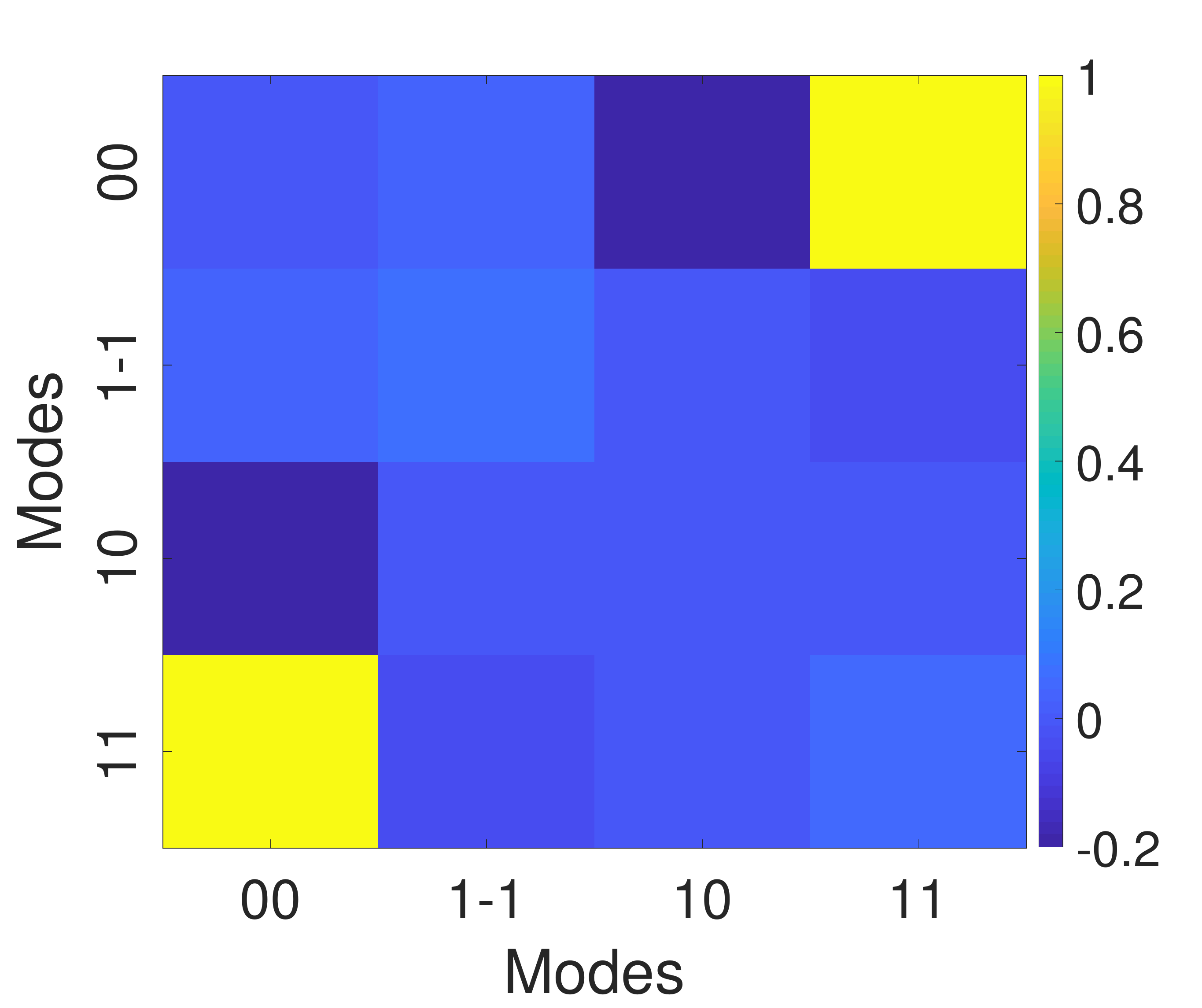}}
  \centerline{{\fontsize{9}{10}\selectfont(d)}}\medskip
\end{minipage}
\caption{Normalized $\left \lvert \fhmc \right\lvert$ at different time instants. The snapshot is taken at $1500$ Hz with a speech source present at (a)-(b) $(\theta, \phi) = (60^\circ, 60^\circ)$ and (c)-(d) $(\theta, \phi) = (60^\circ, 120^\circ)$.}
\label{fig: feature-snapshot}
\end{figure*}
\subsection{Feature selection}
Intuitively, the soundfield coefficients $\alpha_{nm}$ work as natural beamformers in the modal domain due to the inherent properties of the spherical harmonic functions. Hence, the energy distribution of $\alpha_{nm}$ among different modes can be used as a clue for understanding the source directionality. However, there only exists a limited number of active modes in the low frequencies which might prove insufficient to train a neural network for high spatial resolution scenario, especially in a reverberant environment. Therefore, we use the modal coherence model of \eqref{eq: coherence-model} to construct our input feature. For a multi-source scenario, it is common to assume W-disjoint orthogonality \cite{yilmaz2004blind} in the STFT domain, i.e., only a single sound source remains active in each TF bin of the STFT spectrum. Under the W-disjoint orthogonality assumption, \eqref{eq: coherence-model} takes the following form
\begin{multline} \label{eq: coherence-model-single-source-active}
\mathbb{E} \Big\{ \alpha_{nm}(k) \alpha^*_{n'm'}(k) \Big\} = \mathbb{E} \bigg\{ \Big\lvert S_{\ell'}(k) \Big \rvert ^2 \bigg\} \Bigg(  \mathbb{E} \bigg\{ \Big \lvert G_{\ell'}^{(d)}(k) \Big \rvert ^2 \bigg\} \\ \Upsilon_{nm}^{n'm'}(\hat{\boldsymbol{x}}_{\ell'}) +
\sum \limits_{vu}^{V} \mathbb{E} \Big\{ \gamma^{(\ell')}_{vu}(k) \Big\} \text{ } \Psi_{n,n',v}^{m,m',u} \Bigg)
\end{multline}
where $\ell' \in [1, L]$. Note that, \eqref{eq: coherence-model-single-source-active} remains true for a single-source scenario as well.\par
When dealing with audio signals having variable spectral densities, e.g., speech signals, it is intuitive to select a feature based on the relative transfer function to make the feature independent to the variations in the input signal \cite{laufer2017semi}. Following a similar reasoning, we define the relative modal coherence (RMC) as
\begin{multline} \label{eq: relative-modal-coherence-2}
\mathbb{E} \Big\{ \alpha_{nm}(k) \alpha^*_{n'm'}(k) \Big\} \Big/ \mathbb{E} \Big\{ \alpha_{00}(k) \alpha^*_{00}(k) \Big\} \\ 
=\frac{\mathbb{E} \bigg\{ \Big \lvert G_{\ell'}^{(d)}(k) \Big \rvert ^2 \bigg\} \text{ } \Upsilon_{nm}^{n'm'}(\hat{\boldsymbol{x}}_{\ell'}) + \sum \limits_{vu}^{V} \mathbb{E} \Big\{ \gamma^{(\ell')}_{vu}(k) \Big\} \text{ } \Psi_{n,n',v}^{m,m',u}}
{\mathbb{E} \bigg\{ \Big \lvert G_{\ell'}^{(d)}(k) \Big \rvert ^2 \bigg\} \text{ } \Upsilon_{00}^{00}(\hat{\boldsymbol{x}}_{\ell'}) + \sum \limits_{vu}^{V} \mathbb{E} \Big\{ \gamma^{(\ell')}_{vu}(k) \Big\} \text{ } \Psi_{0,0,v}^{0,0,u}} \\
= \frac{\mathbb{E} \bigg\{ \Big \lvert G_{\ell'}^{(d)}(k) \Big \rvert ^2 \bigg\} \text{ } \Upsilon_{nm}^{n'm'}(\hat{\boldsymbol{x}}_{\ell'}) + \sum \limits_{vu}^{V} \mathbb{E} \Big\{ \gamma^{(\ell')}_{vu}(k) \Big\} \text{ } \Psi_{n,n',v}^{m,m',u}}
{4 \pi \text{ } \mathbb{E} \bigg\{ \Big \lvert G_{\ell'}^{(d)}(k) \Big \rvert ^2 \bigg\} + \frac{16 \pi^2}{\sqrt{4 \pi}} \text{ } \mathbb{E} \Big\{ \gamma^{(\ell')}_{00}(k) \Big\}}
\end{multline}
where \eqref{eq: relative-modal-coherence-2} is derived using \eqref{eq: direct-path-upsilon} - \eqref{eq: Wvnn}. From \eqref{eq: relative-modal-coherence-2} it is evident that the relative modal coherence has a direct relation with the source position in a particular room. However, with multiple active sources in a strong reverberant environment, \eqref{eq: relative-modal-coherence-2} introduces additional complexity due to the additive terms in the denominator. On the other hand, the modal coherence of \eqref{eq: coherence-model-single-source-active} offers a simpler alternative to train a CNN due to the fact that the mode-independent term $\bigg\{ \Big\lvert S_{\ell'}(k) \Big \rvert ^2 \bigg\}$ of \eqref{eq: coherence-model-single-source-active} acts merely as a constant scaling factor across different TF bins without altering the relative strength between different modes inside a TF bin. The use of \eqref{eq: coherence-model-single-source-active} as a feature reduces the complexity by eliminating location-dependency from the denominator compared to RMC. Hence, we pose the DOA estimation problem as an image-identification problem for CNN where the feature snapshot is defined as the modal coherence of the soundfield in the individual TF bins of the STFT spectrum
\begin{multline} \label{eq: feature-mc}
    \fhmc(k) = \bigg \{ \mathbb{E} \Big\{ \alpha_{nm}(k) \alpha^*_{n'm'}(k) \Big\} : n \in [0, N], \\
    m \in [-n, n], n' \in [0, N], m' \in [-n', n'] \bigg \}
\end{multline}
where $\fhmc$ is considered as an image consisting of $[\mathcal{N} \times \mathcal{N}]$ complex-valued pixels with $\mathcal{N} = (N+1)^2$ is the total number of modes. Note that, $\fhmc$ is a frequency-dependent function due to the frequency dependency of $\alpha_{nm}$, hence, we need to collect $\fhmc$ from different frequency bands for training so that the model can learn the frequency variations of the feature for the same source position. This deviation is analogous to the transformed image conundrum in an image-identification problem.\par
We also need to train the CNN model for various amplification levels due to the presence of the source PSD term $\mathbb{E} \bigg\{ \Big\lvert S_{\ell'}(k) \Big \rvert ^2 \bigg\}$ in $\fhmc$ (analogous to train a neural network to accommodate the differences in brightness of the same image). This can be achieved through training the CNN model with any non-white random audio signal such that the signal has a variable PSD in both time and frequency directions of the STFT spectrum.\par
Finally, since a CNN model is best suited to work with real data, we convert our $2$D complex-valued feature  $\fhmc$ into corresponding $3$D tensor $\fmc$ of $[\mathcal{N} \times \mathcal{N} \times 2]$ dimension such that
\begin{equation} \label{eq: feature-mc-3d}
\fmc = \Big \langle \Big \langle \mathcal{R}\big\{\fhmc\big\}, \text{ } \mathcal{I}\big\{\fhmc\big\} \Big \rangle \Big \rangle_{3}
\end{equation}
where $ \langle \langle \cdot, \text{ } \cdot \rangle \rangle_{3}$ stacks two matrices in the $3^{rd}$ dimension and $\mathcal{R}\{ \cdot \}$ and $\mathcal{I}\{ \cdot \}$ denote the real and imaginary part, respectively.\par
Fig. \ref{fig: feature-snapshot} shows the normalized snapshots of $\fmc$ captured at random time instants at $1500$ Hz. For a CNN model to work with our input features, we want them to be time-independent for the same source position in a room irrespective of the nature of the audio signal. Indeed, as we observe from Fig. \ref{fig: feature-snapshot}, $\fmc$ changes as a function of source angle and remains fairly constant across time.
\subsection{TF bin processing} \label{sec: tf-bin-precessing}
During both training and evaluation phases, the proposed CNN framework processes each TF bin independently, i.e., it learns the directional patterns based on the spatial distribution of the TF bin energy. Hence, it is important to consider only the TF bins with a significant energy level to avoid misleading the neural network. However, due to the sparse nature of speech signals in both time and frequency, a large proportion of the TF bins usually ends up having low energy. The sparsity in time can be addressed with a suitably designed voice activity detector, but the sparsity along the frequency can still mislead the CNN. Hence, to exclude the low-energy TF bins from the training and evaluation datasets, a energy-based pre-selection of TF bins is required where we drop all the TF bins with an average energy below a certain threshold. If $\mathcal{T}_{\text{all}}$ is denoted as the collection of all the TF bins, we can define a new set $\mathcal{T}_{\text{act}} \subseteq \mathcal{T}_{\text{all}}$ such that
\begin{equation} \label{eq: tf-bin-energy}
    \mathcal{T}_{\text{act}} = \{ \kappa \in \mathcal{T}_{\text{all}} : E_{\kappa} \geq E_{\min} \}
\end{equation}
where $E_{\kappa}$ is the average energy of the spatial coherence matrix for the $\kappa^{th}$ TF bin and $E_{\min}$ is the minimum energy threshold. This lowest energy threshold can be a preset based on empirical measurements, or it can be set dynamically based on the average energy of all the TF bins in the processing block. However, the average energy of the processing block can be low when the number of low-energy TF bins is high. We can also set the minimum energy level at the $\mathcal{K}^{th}$ percentile of the average energy distribution of all TF bins, where $\mathcal{K}$ is usually large for speech signal, as long as we are able to make at least a high-level prediction about the energy distribution among the TF bins. Note that, \eqref{eq: tf-bin-energy} should be applied at both training and evaluation stage, however, $E_{\min}$ does not need to be the same.\par
A second issue may arise when a TF bin violates the W-disjoint orthogonality principle. In the proposed algorithm, CNN predicts the most dominant source in each TF bin which are later combined in a clustered histogram to reach a global outcome. However, as shown in \cite{yilmaz2004blind}, the number of TF bins violating the W-disjoint orthogonality increases as the number of simultaneous sources increases. When a TF bin contains significant energy from multiple sources, the prediction of the CNN model can be arbitrary. To circumvent the uncertainty in prediction due to the violation of W-disjoint orthogonality principle, we only consider the predictions in the TF bins where the CNN model predicts a single DOA with a high confidence level. Hence, if we define the probability score for each TF bin as
\begin{equation} \label{eq: tf-bin-probabilty-score-el}
    \boldsymbol{\mathcal{P}}_{\kappa, \Theta} = \big \{ \mathcal{P}_{\kappa}(\theta) \big \}_{\theta \in \Theta}
\end{equation}
\begin{equation} \label{eq: tf-bin-probabilty-score-az}
    \boldsymbol{\mathcal{P}}_{\kappa, \Phi} = \big \{ \mathcal{P}_{\kappa}(\phi) \big \}_{\phi \in \Phi}
\end{equation}
where $\mathcal{P}_{\kappa}(\theta)$ and $\mathcal{P}_{\kappa}(\phi)$ are the probability scores of the corresponding elevation and azimuth classes at the $\kappa^{th}$ TF bin, the final DOA estimation at the evaluation stage should be based on the set $\mathcal{T}_{\text{test}} \subseteq \mathcal{T}_{\text{act}}$ such that\par
\begin{multline} \label{eq: tf-bin-probabilty-filter}
    \mathcal{T}_{\text{test}} = \Big \{ \kappa \in \mathcal{T}_{\text{act}} : \max \big \{ \boldsymbol{\mathcal{P}}_{\kappa, \Theta} \big \} \geq \mathcal{P}_{\min} \text{ and } \\
    \max \big \{ \boldsymbol{\mathcal{P}}_{\kappa, \Phi} \big \} \geq \mathcal{P}_{\min} \Big \}
\end{multline}
where $\mathcal{P}_{\min}$ denotes the minimum confidence level.
\subsection{CNN Architecture} \label{sec: cnn-arch}
We utilize a CNN to estimate source DOA based on the local connectivity of the modal coherence coefficients. A CNN topology typically consists of multiple convolution layers followed by fully-connected networks. For DOA estimation, we perform multi-output multi-class classification where we share the same convolution layer structure to predict both the azimuth and elevation using separate fully connected heads at the last stage - each responsible for predicting either azimuth or elevation. We opt for a classification-based approach due to the limited resolution of the practical dataset. However, the proposed technique can be studied with a regression-based model subject to the availability of a denser training grid to learn the evolution of dynamic reverberation characteristics.\par
In each convolutional layer, we use $64$ spatial filters of $2 \times 2$ size to learn the spatial coherence pattern for each desired point in a predefined DOA grid. As our feature is defined as the modal coherence for each TF bin, it is important to consider $2$D filters in the convolution layers. A rectified linear unit (ReLU) activation follows the convolution layer at each stage. The evaluation is done with $8$ convolution layers with zero padding to keep the output size same for each layer. The final convolution layer is connected to $2$ fully-connected layers that use ReLU activation. Finally, two separate fully connected heads responsible for azimuth and elevation estimation, respectively, are used with Sigmoid activation. A Sigmoid activation is chosen over Softmax in the last stage as it allows us to perform prediction-based TF bin selection to remove the bins with low confidence, as described in Section \ref{sec: tf-bin-precessing}.\par
Due to the W-disjoint orthogonality assumption, ideally each TF bin is designated with a single DOA and can be classified using a multi-class classification network using a Softmax activation-based categorical cross-entropy loss. However, in a practical environment with multiple simultaneously active sources, it is unrealistic to expect each of the TF bins to honor the W-disjoint orthogonality. Therefore, it is possible to find occasional TF bins whose energy are contributed by multiple sound sources. Such a TF bin produces a feature snapshot which does not match with any of the patterns learned by the model during the single-source training stage. In such cases it is expected that the output will not have a large prediction score for any of the classes. Hence, we use binary cross-entropy loss function with Sigmoid activation instead of categorical cross-entropy in order to independently predict the probability of each individual class in every TF bin. This approach allows us to enforce the criterion mentioned in \eqref{eq: tf-bin-probabilty-filter}.\par
Detailed parameter settings are included in the experimental results section.
\subsection{Training the model}
In the training stage, we train the model based on the feature snapshots in a single source scenario. Each training data is labeled independently for azimuths and elevations. The model is trained for the elevation set $\Theta$ and the azimuth set $\Phi$ in different azimuth and elevation planes, respectively. Once the model learns the patterns for each of the intended directions, we independently predict the elevations and azimuths for any number of concurrent sources as long as the W-disjoint orthogonality principle majorly holds. This is a more realistic approach than training the model for each possible angular combination \cite{adavanne2018direction,chakrabarty2019multi} which becomes a resource-intensive operation as the number of classes or the number of simultaneous sources increases. Furthermore, the proposed method does not require retraining the model every time an additional source appears in the mixture.
\begin{algorithm}[!t] \label{algo: doa-training}
  \SetAlgoLined
  \KwData{$\Theta, \Phi, \alpha_{nm}(\theta, \phi) \text{ } \forall nm \text{, } \forall \theta \in \Theta \text{, } \forall \phi \in \Phi$}
  Calculate spatial coherence $\mathbb{E} \Big\{ \alpha_{nm}(k) \alpha^*_{n'm'}(k) \Big\}$ in each TF bin using \eqref{eq: expected-value-mc};\\
  Get $\fhmc \text{ } \forall \theta \in \Theta \text{, } \forall \phi \in \Phi$  using \eqref{eq: feature-mc};\\
  Apply \eqref{eq: tf-bin-energy} to filter out low energy TF bins and get $\mathcal{T}_{\text{act}}$;\\
  Use $\mathcal{T}_{\text{act}}$ to train the model using the parameters in Table \ref{table: evaluation-parameters} independently for $\Theta$ and $\Phi$;\\
  Save the model
  \caption{Algorithm for DOA estimation - training stage}
\end{algorithm}
\begin{algorithm}[!t] \label{algo: doa-test}
  \SetAlgoLined
  \KwData{$\alpha_{nm} \text{ } \forall nm$}
  Calculate spatial coherence $\mathbb{E} \Big\{ \alpha_{nm}(k) \alpha^*_{n'm'}(k) \Big\}$ in each TF bin using \eqref{eq: expected-value-mc};\\
  Get $\fhmc$  using \eqref{eq: feature-mc};\\
  Apply \eqref{eq: tf-bin-energy} to filter out low energy TF bins and get $\mathcal{T}_{\text{act}}$;\\
  Calculate the probability of each classes in $\Theta$ and $\Phi$ for the TF bins in $\mathcal{T}_{\text{act}}$ using the model saved during training;\\
  Apply \eqref{eq: tf-bin-probabilty-filter} to get $\mathcal{T}_{\text{test}}$;\\
  Apply \eqref{eq: defintion-prediction-tf} to form the prediction multiset $\mathcal{X}$;\\
  \eIf{$L==1$}{
       Use \eqref{eq: final-output-no-clustering} to estimate DOA;
   }{
       Using a suitable clustering algorithm, divide $\mathcal{X}$ into $L$ clusters;\\
       Use \eqref{eq: final-output-clustering} to estimate $L$ source directions.
  }
  \caption{Algorithm for DOA estimation - evaluation stage}
\end{algorithm}
\subsection{DOA estimation} \label{sec: doa-estimation}
First, we jointly pick the highest probable elevation and azimuth classes for each TF bin in $\mathcal{T}_{\text{test}}$ to form prediction multiset $\mathcal{X}$
\begin{multline} \label{eq: defintion-prediction-tf}
    \mathcal{X} = \Big \{ \Big( \argmax_{\theta \in \Theta} \big \{ f : \theta \mapsto \mathcal{P}_{\kappa}(\theta) \big \} \text{, } \\
    \argmax_{\phi \in \Phi} \big \{ f : \phi \mapsto \mathcal{P}_{\kappa}(\phi) \big \} \Big) : \kappa \in \mathcal{T}_{\text{test}} \Big \}.
\end{multline}
Subsequently, the simplest way of multi-source DOA estimation is to pick $L$ largest peaks from the $2$D histogram of $\mathcal{X}$, i.e.,
\begin{equation} \label{eq: final-output-no-clustering}
    \big \{ \doublehat{\boldsymbol{x}} \big \} = \big \{ (\theta, \phi) \text{ of $L$ largest peaks in } \mathcal{X} \big \}.
\end{equation}
\par
However, in case of a noisy prediction for a multi-source environment, the aforementioned technique can cause erroneous results. For example, in a $2$-source environment, if the true DOA of the prominent source lies between two adjacent classes, both the adjacent classes for the prominent source might occur more frequently than the true class corresponding to the weaker source. To avoid such a scenario, a more robust technique is to apply a suitable clustering algorithm, such as k-means \cite{loesch2008source} or density-based \cite{hafezi2018robust} clustering, to divide $\mathcal{X}$ into $L$ clusters and pick the peak in each cluster, i.e.,
\begin{equation} \label{eq: final-output-clustering}
    \big \{ \doublehat{\boldsymbol{x}} \big \}  = \big \{ (\theta, \phi) \text{ of the peak of $L$ clusters in } \mathcal{X} \big \}.
\end{equation}
\par
The training and evaluation steps are outlined in Algorithm \ref{algo: doa-training} and \ref{algo: doa-test}, respectively.
\section{Experimental results and discussion} \label{sec: experimental-results}
In this section, we present the experimental results, comparison, and discussion of the proposed algorithm with the contemporary counterparts.
\subsection{Experimental methodology} \label{sec: experimental-methodology}
We evaluated the proposed method in simulated and practical environments under different room conditions. The parameter settings used in the evaluation are listed in Table \ref{table: evaluation-parameters}. We assessed the performance of the model in $3$ simulated room environments (room S1, S2, and S3 in Table \ref{table: test-environments}) generated using a RIR Generator \cite{habets2006room} as well as with the recordings from a practical room (room P1 in Table \ref{table: test-environments}) in the presence of babble noise. The training and the tests were performed under the same room environment. The reverberation time ($T_{60}$) and direct to reverberation ratio (DRR) shown in Table \ref{table: test-environments} for room P1 were calculated using the techniques outlined in \cite{eaton2015ace}. We only considered the first-order harmonic coefficients which need at least $4$ microphones to calculate, however, in the result section, we used recordings from $9$ microphones oriented on a spherical grid suggested in \cite{fliege1996two} for evaluating the proposed as well as the competing methods.\par
\begin{table}[!t]
\caption{Experimental parameter settings}
\label{table: evaluation-parameters}
\centering
\renewcommand{\arraystretch}{1.2}
\begin{tabular}{@{}ll@{}}
\toprule
Name & Value \\
\midrule
\multicolumn{2}{@{}l@{}}{\textbf{Model parameters}} \\
$\qquad \mathcal{P}_{\min}$ & 0.5 \\
$\qquad E_{\min}$ & Percentile-based \\
$\qquad \mathcal{K}$ & $90$ \\
$\qquad N$ & 1 \\
\multicolumn{2}{@{}l@{}}{\textbf{CNN parameters}} \\
$\qquad$Input size & $[4 \times 4 \times 2]$ \\
$\qquad$\# Convolution layers & $8$ \\
$\qquad$\# Conv. filters & $64$ $([2 \times 2])$ \\
$\qquad$\# Dense layers & $2$ $(512)$ \\
\bottomrule
\end{tabular}
\end{table}
We used random mixed-gender speech signals from the TIMIT corpus \cite{garofolo1993timit} to synthesize the reverberant signals from the measured/simulated RIRs. The spherical harmonic decomposition, as described in Section \ref{sec: spherical-harmonics-decomposition}, was performed on the reverberant signals to calculate the spherical harmonic coefficients up to first order. Note that, the proposed method is independent of the type and shape of the sensor array as long as the array is capable of performing spherical harmonic decomposition\footnote{A number of alternate array structures are available in the literature for capturing spherical harmonic coefficients, a few can be found in \cite{abhayapala2002theory,li2007flexible,chen2015theory,meyer2002highly,samarasinghe2017planar,abhayapala2010spherical}.}.\par
The CNN architecture has been discussed and presented in Section \ref{sec: cnn-arch} and Table \ref{table: evaluation-parameters}. The implementation was done in Python using Keras \cite{chollet2015keras} running on top of TensorFlow \cite{tensorflow2015-whitepaper}. For the proposed method, the TF bin-level predictions were accumulated and clustered using k-means algorithm (step 10 in Algorithm \ref{algo: doa-test}) assuming that the number of active sources was known as \textit{a priori}, however, certain algorithms offer to cluster the data without the advance knowledge of the number of sources \cite{hafezi2018robust}. Furthermore, as k-means algorithm works with the Euclidean geometry, we converted the predicted DOAs to corresponding Cartesian coordinates on a unit sphere before clustering the data.\par
The processing was done at $16$ kHz sampling frequency. The STFT used a $16$ ms Hanning window, $50\%$ overlap, and $256$-point discrete Fourier transform (DFT). We only utilized the frequencies ranged $500-2000$ Hz for DOA estimation. A $30$s long speech was used to synthesize the data for training, however, the actual number of features reduced significantly after applying \eqref{eq: tf-bin-energy} to filter out low-energy TF bins. The majority of the results and discussions in this section are presented for azimuth estimation only considering the fact that the estimation of evaluation is independent of the azimuth estimation and follows the same mechanism. However, in Section \ref{sec: result-joint-estimation}, we have demonstrated how a joint azimuth and elevation estimation can be performed using the proposed method.
\begin{table}[!t]
\caption{Test environments. $d_{sm}$ denotes source to microphone distance.}
\label{table: test-environments}
\centering
\renewcommand{\arraystretch}{1.2}
\begin{tabular}{@{}lcccc@{}}
\toprule
Room & Dimension & $T_{60}$ & DRR & $d_{sm}$ \\
\midrule
P1 & $[11 \times 7.5 \times 2.75]$ m & $640$ ms & $-0.6$ dB & $2.8$ m \\
S1 & $[6 \times 4 \times 3]$ m & $200$ ms & - & $1$ m \\
S2 & $[7 \times 6 \times 3]$ m & $300$ ms & - & $1$ m \\
S3 & $[8 \times 6 \times 3]$ m & $500$ ms & - & $1$ m \\
\bottomrule
\end{tabular}
\end{table}
\subsection{Baseline methods and evaluation metrics}
The performance of the proposed algorithm is compared with a recent CNN-based DOA estimation method proposed in \cite{chakrabarty2019multi} (subsequently denoted as \say{CNN-PH}) where it was already shown that \say{CNN-PH} outperforms conventional parametric methods like MUSIC and SRP-PHAT. For a fair comparison, we kept the CNN architecture and other evaluation criteria same in all possible ways. We used the same $9$-microphone setup as described in Section \ref{sec: experimental-methodology} for the competing methods unless mentioned otherwise. The convolution filter size for \say{CNN-PH} was set to $[2 \times 1]$ as per the recommendation of the authors \cite{chakrabarty2019multi} whereas we applied $[2 \times 2]$ filters with the proposed method. The difference in filter size between the competing methods comes from the fact that the feature used in \say{CNN-PH} spans across the frequency band where multiple active sources can be present in the horizontal dimension. On the other hand, the proposed method uses the modal coherence snapshot of a single TF bin as a feature where only one active source is expected due to the assumption of W-disjoint orthogonality.\par
\begin{figure*}[!t]
\hspace{9mm}
\begin{minipage}[b]{0.89\linewidth}
  \centering
  \centerline{\includegraphics[width=\linewidth]{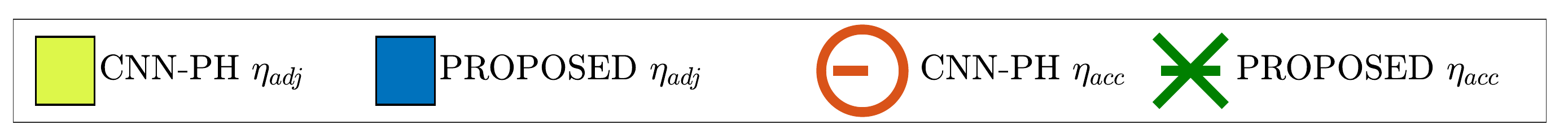}}
\end{minipage}\\
\centering
\begin{minipage}[b]{0.30\linewidth}
  \centering
  \centerline{\includegraphics[width=\linewidth]{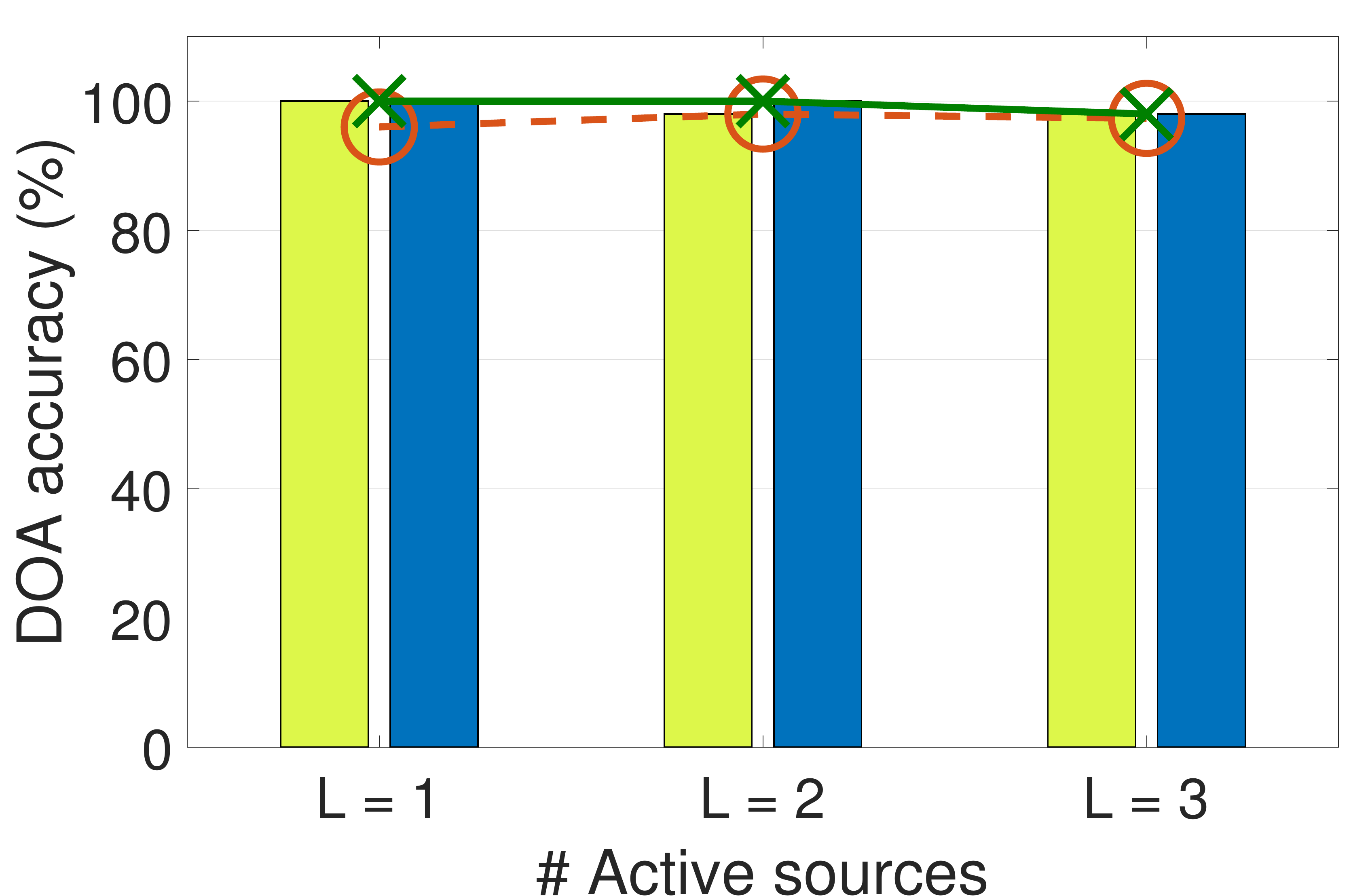}}
  \centerline{{\fontsize{9}{10}\selectfont(a) Room S1, $30$dB SNR}}\medskip
\end{minipage}
\begin{minipage}[b]{0.30\linewidth}
  \centering
  \centerline{\includegraphics[width=\linewidth]{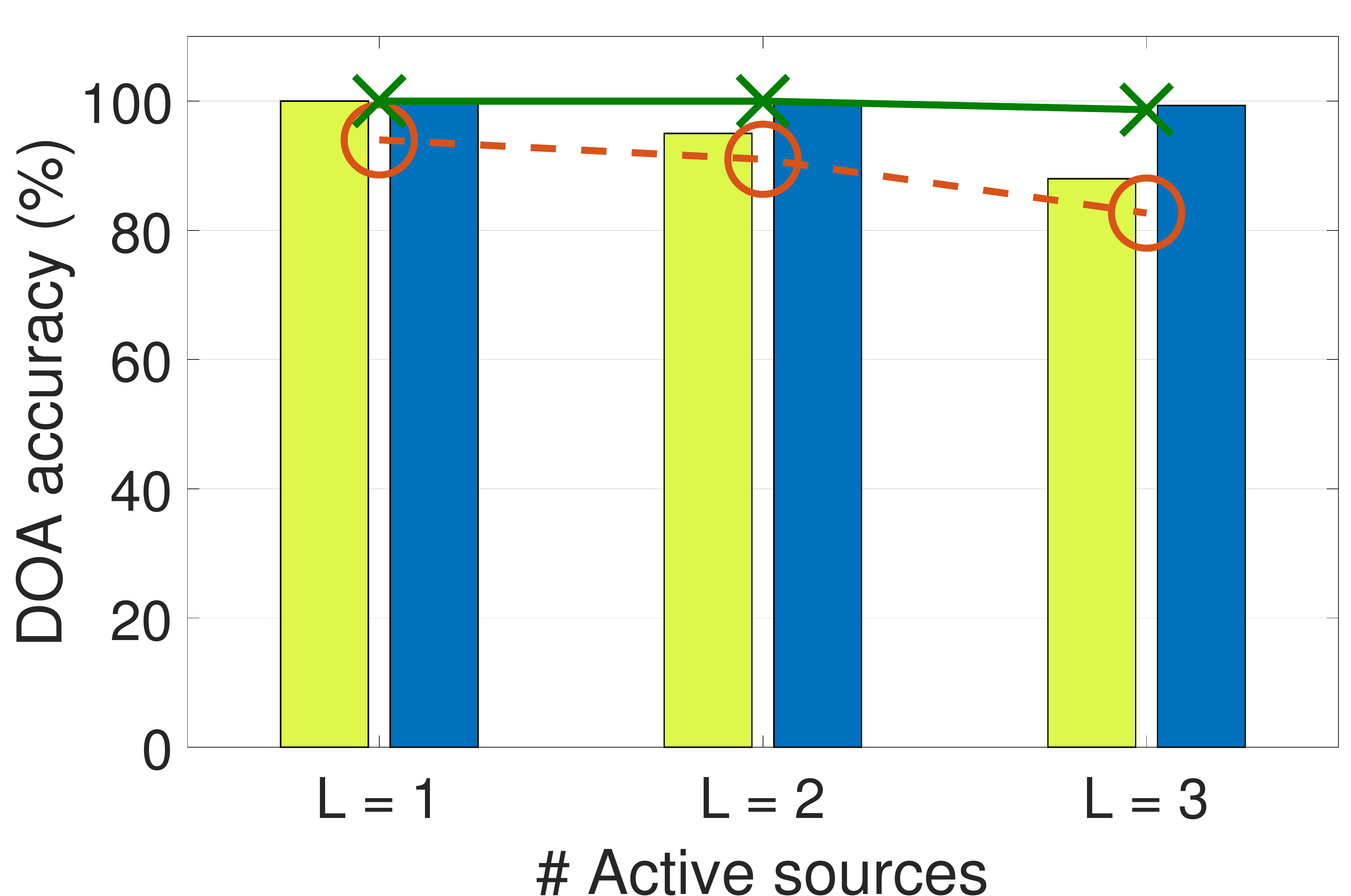}}
  \centerline{{\fontsize{9}{10}\selectfont(b) Room S1, $20$dB SNR}}\medskip
\end{minipage}
\begin{minipage}[b]{0.30\linewidth}
  \centering
  \centerline{\includegraphics[width=\linewidth]{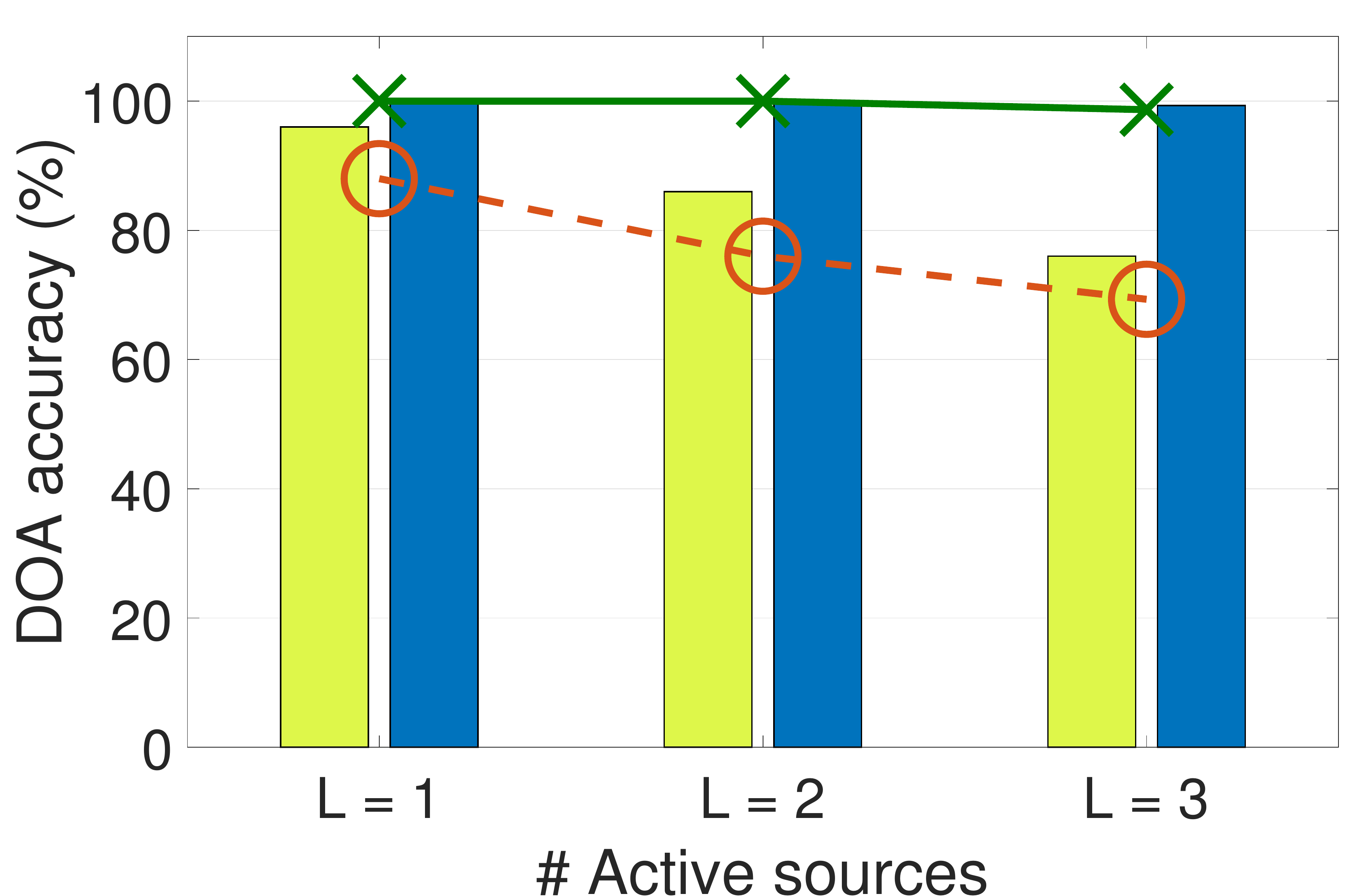}}
  \centerline{{\fontsize{9}{10}\selectfont(b) Room S1, $10$dB SNR}}\medskip
\end{minipage}\\
\begin{minipage}[b]{0.30\linewidth}
  \centering
  \centerline{\includegraphics[width=\linewidth]{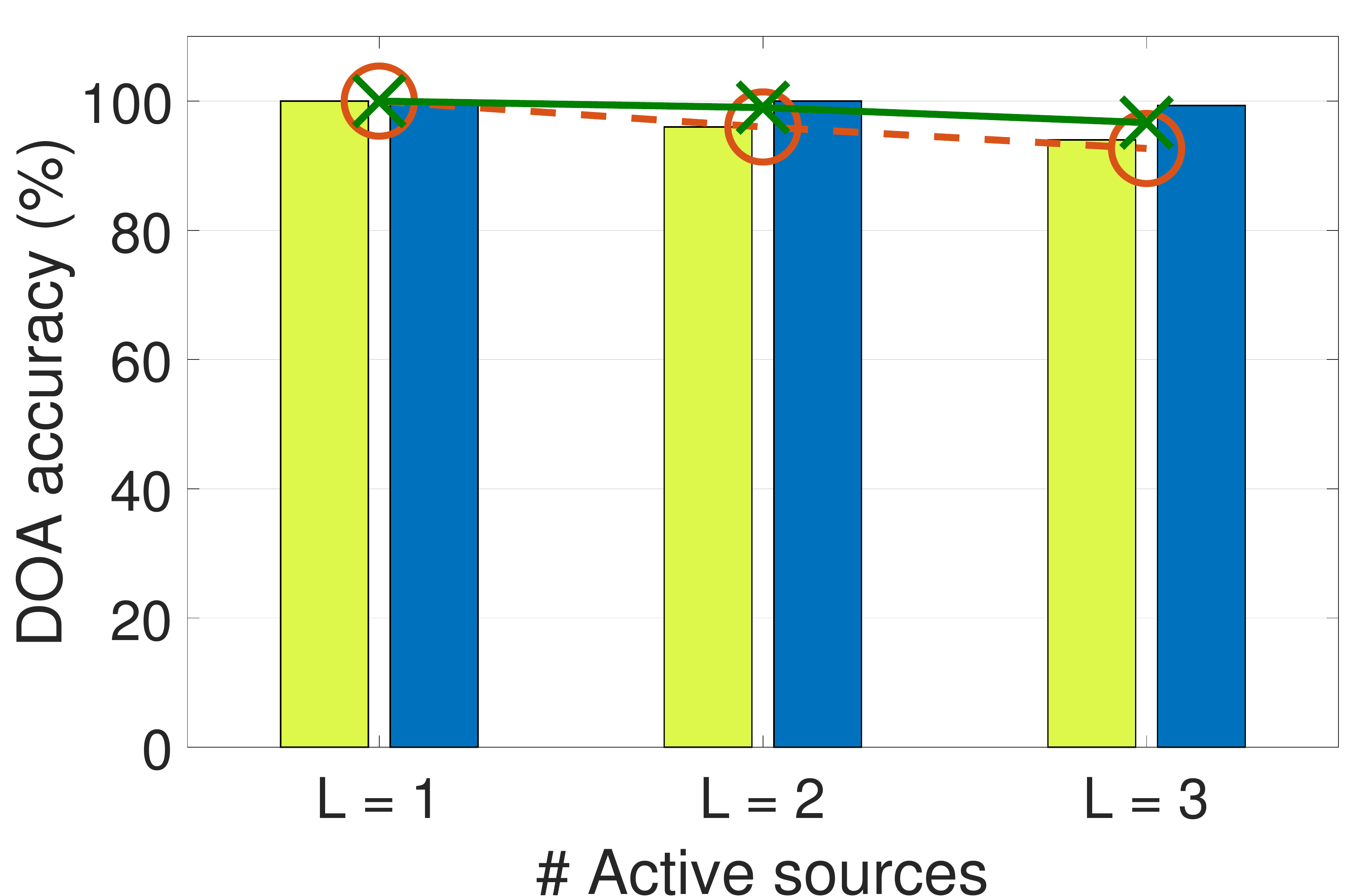}}
  \centerline{{\fontsize{9}{10}\selectfont(c) Room S2, $30$dB SNR}}\medskip
\end{minipage}
\begin{minipage}[b]{0.30\linewidth}
  \centering
  \centerline{\includegraphics[width=\linewidth]{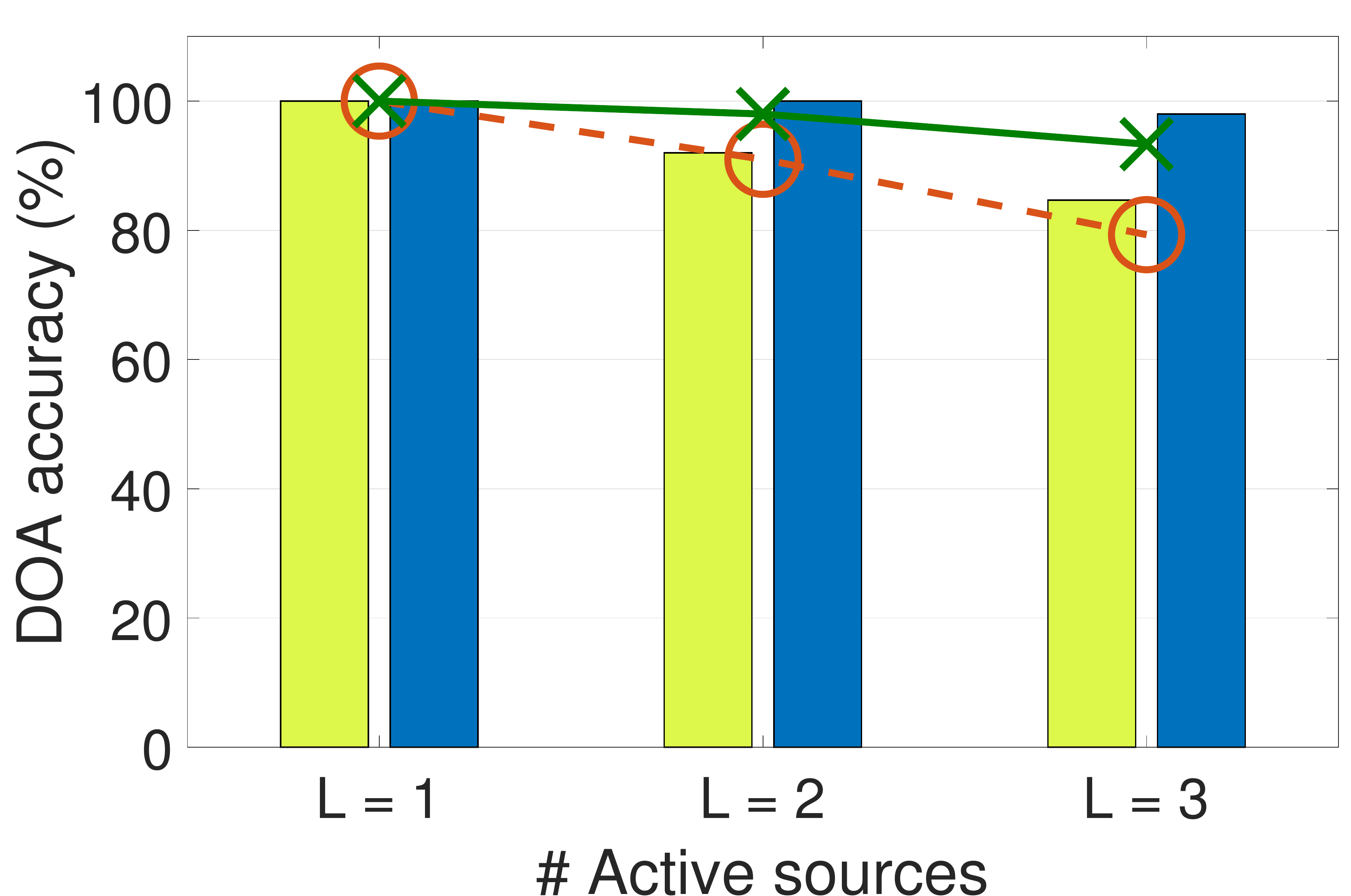}}
  \centerline{{\fontsize{9}{10}\selectfont(d) Room S2, $20$dB SNR}}\medskip
\end{minipage}
\begin{minipage}[b]{0.30\linewidth}
  \centering
  \centerline{\includegraphics[width=\linewidth]{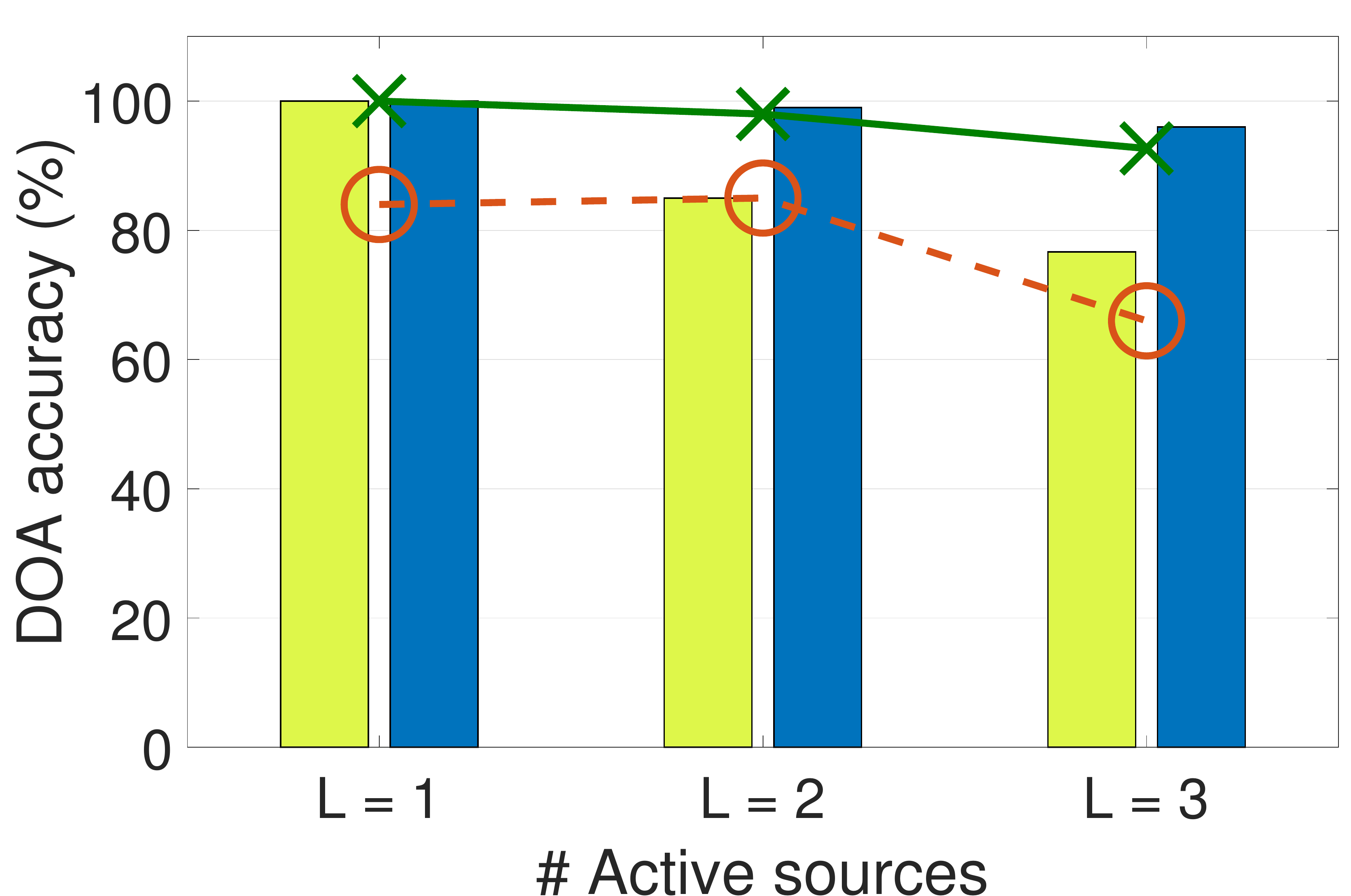}}
  \centerline{{\fontsize{9}{10}\selectfont(d) Room S2, $10$dB SNR}}\medskip
\end{minipage}\\
\begin{minipage}[b]{0.30\linewidth}
  \centering
  \centerline{\includegraphics[width=\linewidth]{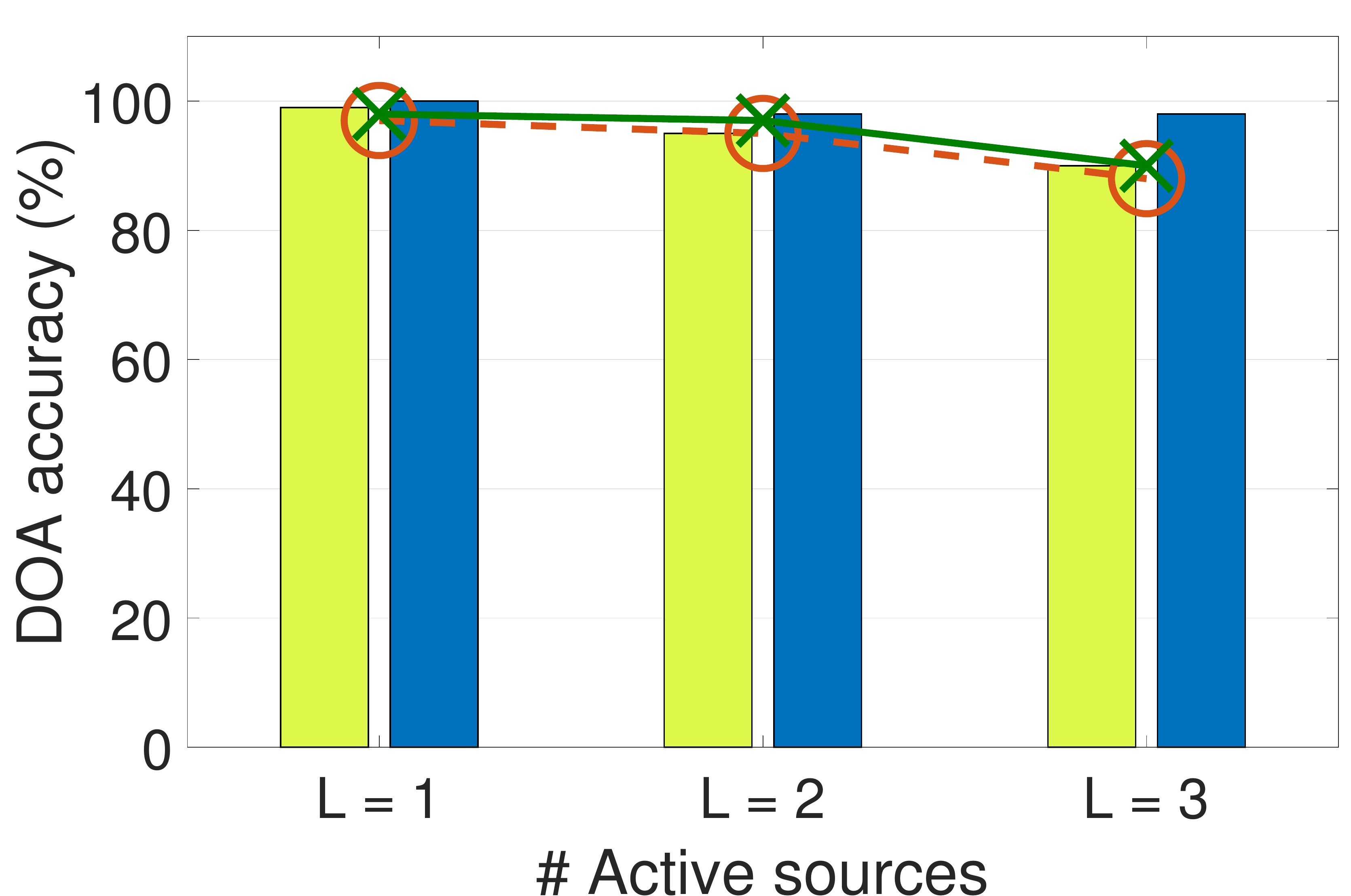}}
  \centerline{{\fontsize{9}{10}\selectfont(e) Room S3, $30$dB SNR}}\medskip
\end{minipage}
\begin{minipage}[b]{0.30\linewidth}
  \centering
  \centerline{\includegraphics[width=\linewidth]{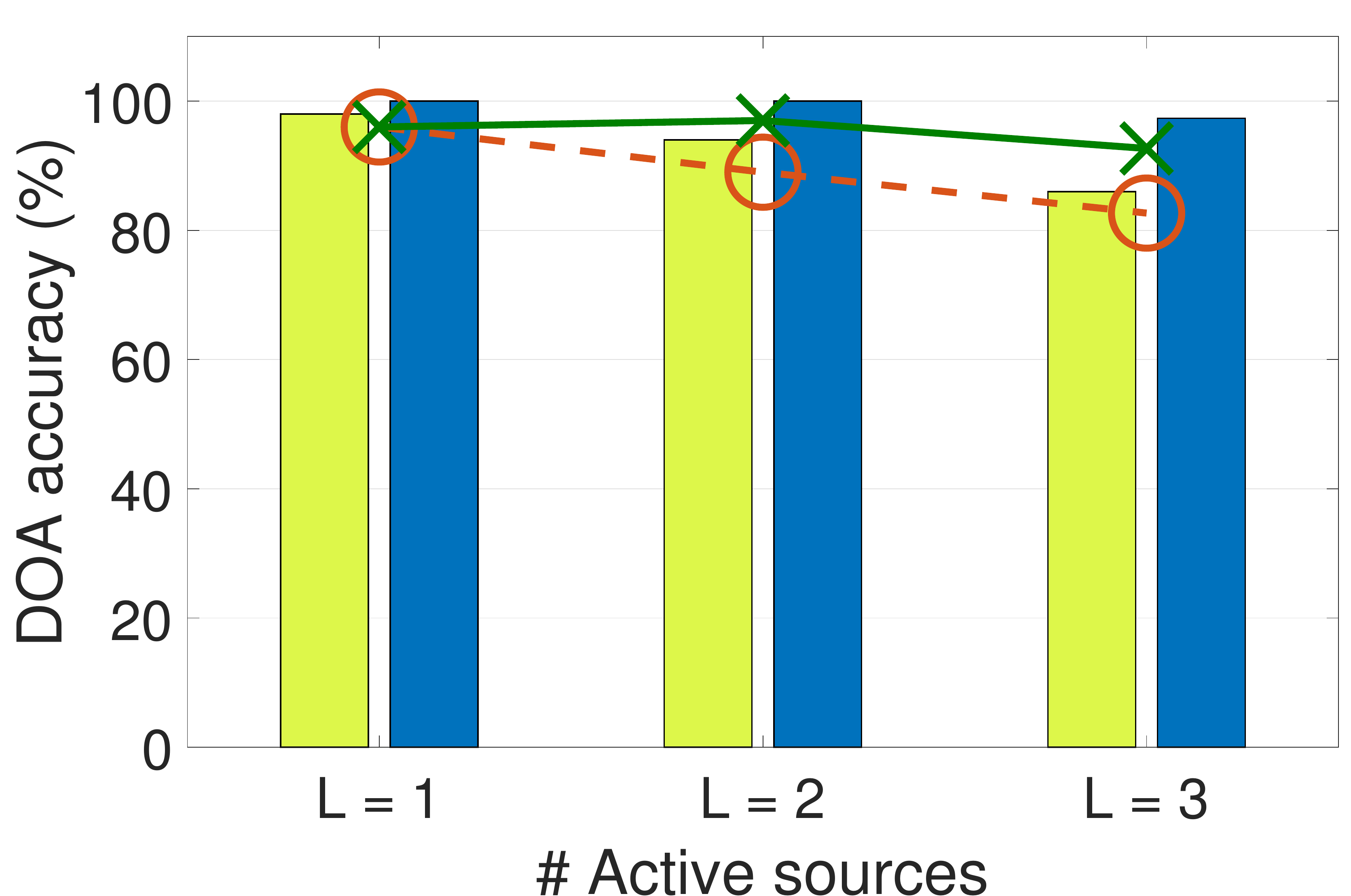}}
  \centerline{{\fontsize{9}{10}\selectfont(f) Room S3, $20$dB SNR}}\medskip
\end{minipage}
\begin{minipage}[b]{0.30\linewidth}
  \centering
  \centerline{\includegraphics[width=\linewidth]{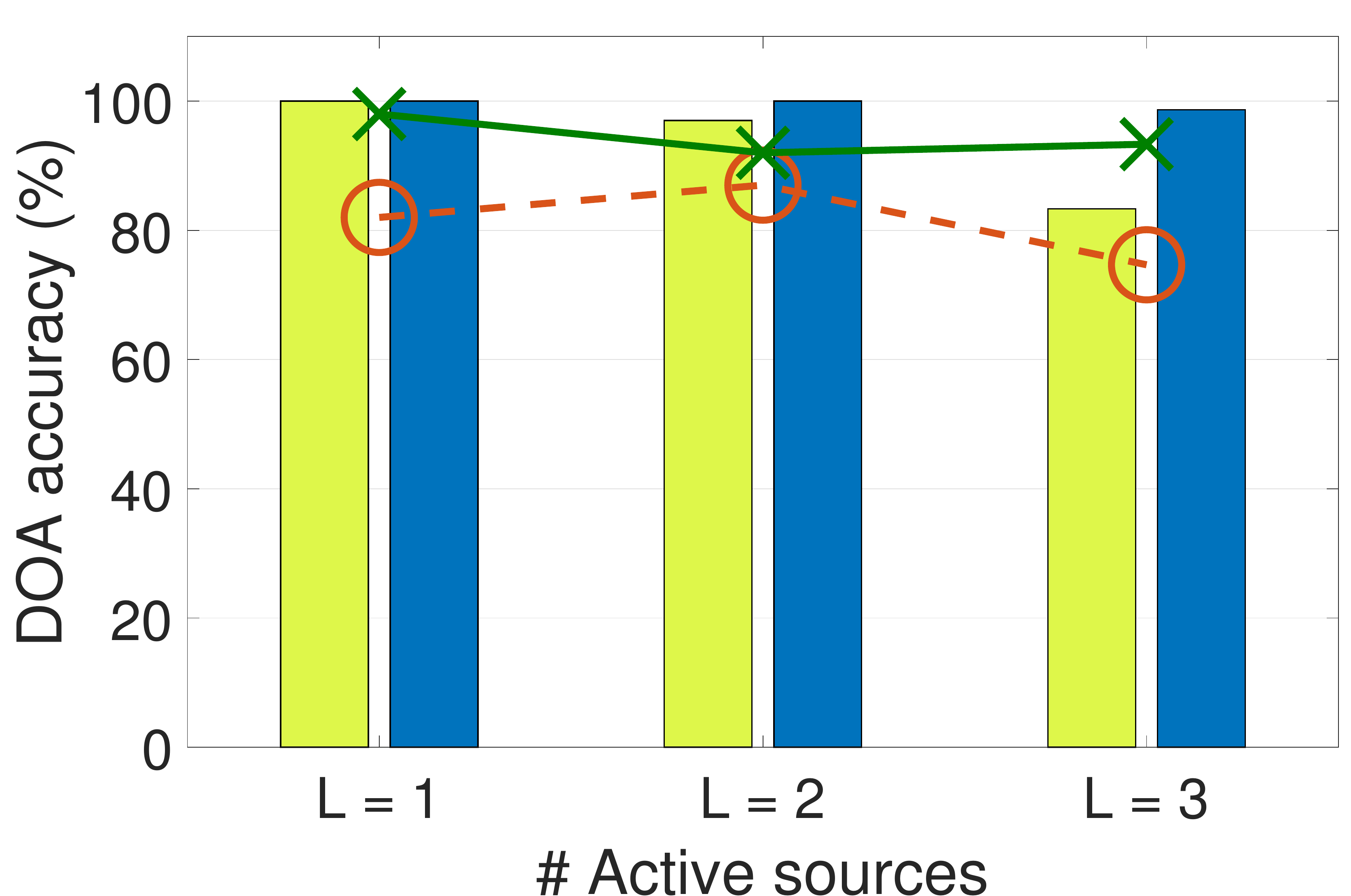}}
  \centerline{{\fontsize{9}{10}\selectfont(f) Room S3, $10$dB SNR}}\medskip
\end{minipage}
\caption{Azimuth estimation under different simulated reverberant and noisy environments on a $45^\circ$ elevation plane.}
\label{fig: result-same-plane}
\end{figure*}
To evaluate the performance, we first defined the prediction error for the $\ell^{th}$ source in a single test by the angular difference between the true and the estimated points at the origin of a unit sphere, i.e.,
\begin{equation} \label{eq: angular-difference}
    \Delta_{\ell}  = \cos^{-1} \Big [ \cos(\hat{\theta}_{\ell}) \cos(\theta_{\ell}) + \sin(\hat{\theta}_{\ell}) \sin(\theta_{\ell}) \cos(\hat{\phi}_{\ell} - \phi_{\ell}) \Big ].
\end{equation}
As we are posing the DOA estimation as a classification problem, the mean error can be misleading unless the angular difference between adjacent classes are very small. Hence, instead we propose to use performance metrics based on estimation accuracy. At first, we define the multi-source DOA classification accuracy as the percentage of the correct predictions, i.e.,
\begin{equation} \label{eq: accuracy}
    \eta_{\text{acc}} = \frac { \mathcal{M}_{\{\{\Delta_{\ell} \} \} }(0) } { \overline{ \overline{ \{\{\Delta_{\ell} \}\} }} } \times 100 \%
\end{equation}
where $\mathcal{M}_{\mathcal{X}} ( \vartheta )$ denotes the multiplicity of $\vartheta$ in the multiset $\mathcal{X}$, $\{\{ \Delta_{\ell} \}\}$ is a multiset containing $\Delta_{\ell} \text{ } \forall \ell$ for all the tests, and $\overline{ \overline{ \text{ } \cdot \text{ }  }}$ denotes the cardinality of the underlying multiset. Note that, for a single test, the sequence of the true and estimated DOAs need not be in the same order, hence, we map them in such a way that $\mathcal{M}_{\{\{\Delta_{\ell} \} \} }(0)$ is maximized.\par
Occasionally, the definition of \eqref{eq: accuracy} may fail to offer the full picture as it does not take it into consideration how far a wrong prediction deviates from the true value although the adjacent classes are highly correlated in a DOA classification task. Hence, we define another accuracy metric, termed as adjacent accuracy, where we consider the predictions for adjacent classes as true positives as well, i.e.,
\begin{equation} \label{eq: accuracy-2}
    \eta_{\text{adj}} = \frac { \mathcal{M}_{\{\{ \max [0, \Delta_{\ell} - \Delta_{\Omega} ] \} \} }(0) } { \overline{ \overline{ \{\{\Delta_{\ell} \}\} }} } \times 100 \%
\end{equation}
where $\Delta_{\Omega}$ is the angular separation between two adjacent classes. A high $\eta_{\text{adj}}$ with low $\eta_{\text{acc}}$ indicates that the transition of the feature pattern is not very sharp between the adjacent classes, a phenomenon expected in a noisy environment.\par
All the results presented in the subsequent sections are based on the accumulation of the results of $50$ random experiments in each test case. Each experiment was evaluated with random source positions and subsequently added with random Gaussian noise unless specified otherwise.
\subsection{Results and discussions} \label{sec: results-discussions}
In this section, we discuss and compare DOA estimation performances under different criteria and room environments. During the experiments, the microphone array was placed at the center of the room at $1$ m height. For the proposed method, we completed the training once per room considering a single-source scenario and used the same trained model at the testing stage irrespective of the number of simultaneously active sources. A $30$s long speech data were synthesized during the training stage, however, we only used the top $10\%$ STFT bins based on TF bin energy to train the network\footnote{As an instance, the average size of the training dataset in Section \ref{sec: results-sim} was $10,505$ samples per class, each being a $[4 \times 4 \times 2]$ tensor.}.\par
\subsubsection{\textbf{Azimuth estimation for a fixed elevation}} \label{sec: results-sim}
The first set of experiments considered the same elevation plane at $45^\circ$ for both training and testing. We considered uniformly spaced azimuth points at $10^\circ$ interval (i.e., $J = 36$) which makes the angular separation between adjacent classes $\Delta_\Omega = 7.07^\circ$ on the $45^\circ$ elevation plane. For each room, we emulated $2$ different signal to noise ratio (SNR) by adding white Gaussian noise and evaluated the performance for up to $3$ active sources, i.e., $L = [1, 3]$. As \say{CNN-PH} was originally designed to be trained for all possible angular combinations in an $L$-source DOA estimation, we trained \say{CNN-PH} for $36$ and $1260$ unique angular combinations based on $36$ azimuth classes for $L=1$ and $2$, respectively\footnote{The training process for \say{CNN-PH} is outlined in \cite[pp. 13]{chakrabarty2019multi}}. However, for testing with $L = 3$, we trained \say{CNN-PH} for $2$-source mixture ($1260$ angular combinations) to understand the performance in a dynamic acoustic scenario. In contrast, the proposed method was always trained for the single-source scenario, e.g., $36$ unique cases for this experiment, irrespective of the number of sources in the testing environment.\par
\begin{figure}[!t]
\centering
\begin{minipage}[b]{0.80\linewidth}
  \centering
  \centerline{\includegraphics[width=\linewidth]{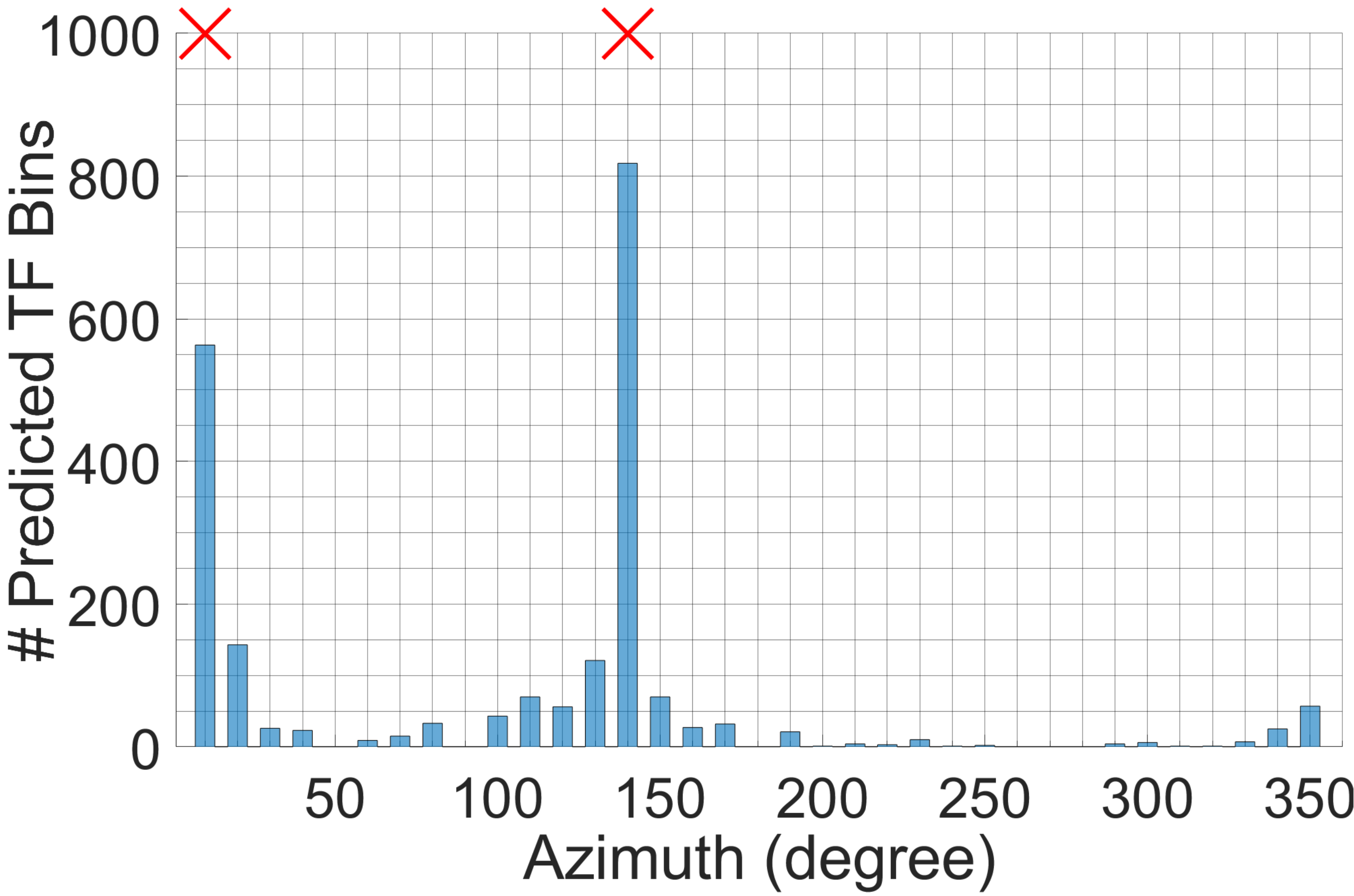}}
  \centerline{{\fontsize{9}{10}\selectfont(a)  $ L = 2$}}\medskip
\end{minipage}
\begin{minipage}[b]{0.80\linewidth}
  \centering
  \centerline{\includegraphics[width=\linewidth]{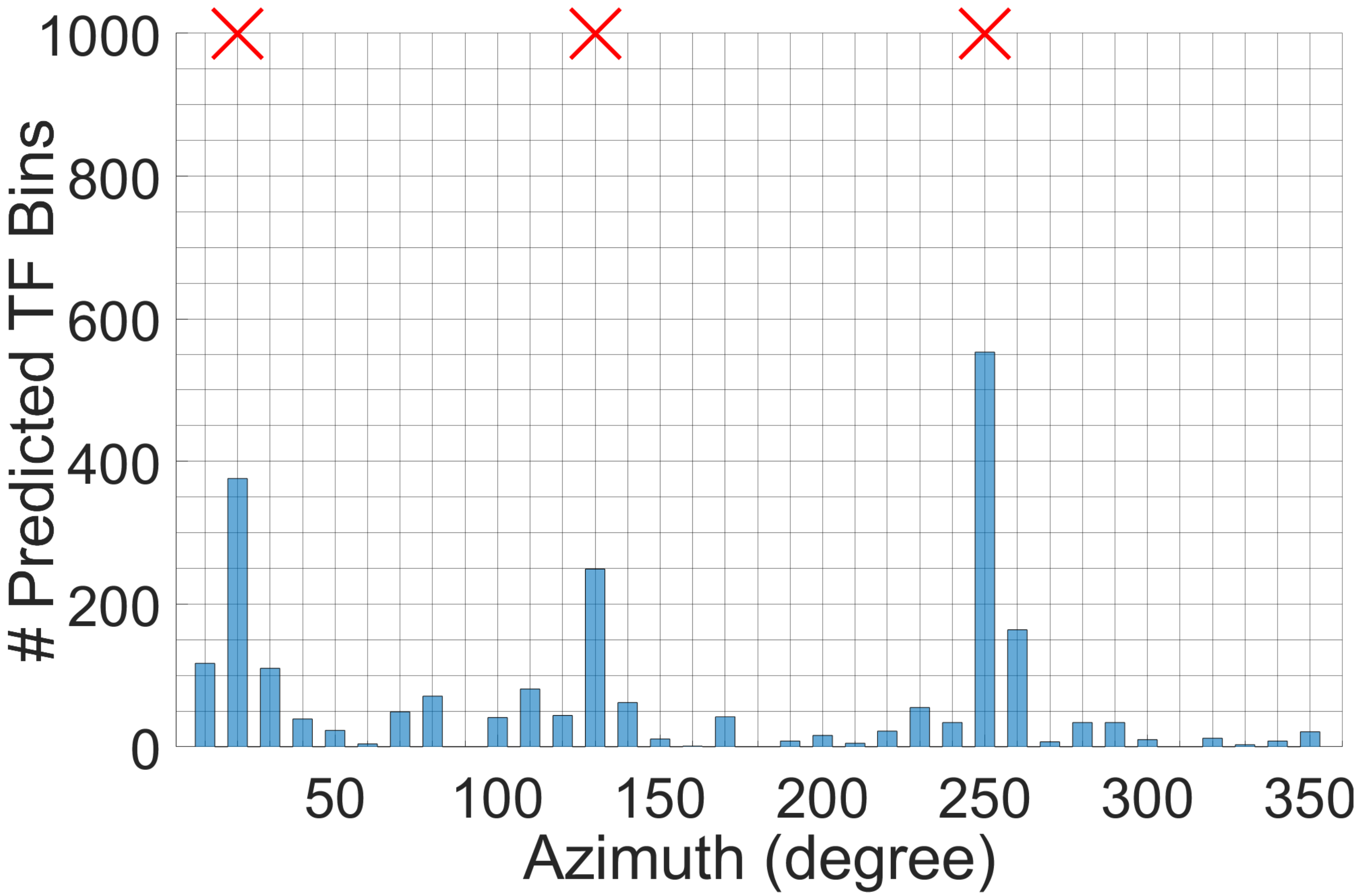}}
  \centerline{{\fontsize{9}{10}\selectfont(b)  $ L = 3$}}\medskip
\end{minipage}
\caption{TF bin prediction histogram in room S3 ($T_{60} = 500$ ms). Red crosses denote the ground truths.}
\label{fig: histogram-1}
\end{figure}
\begin{figure}[!t]
\centering
\begin{minipage}[b]{0.80\linewidth}
  \centering
  \centerline{\includegraphics[width=\linewidth]{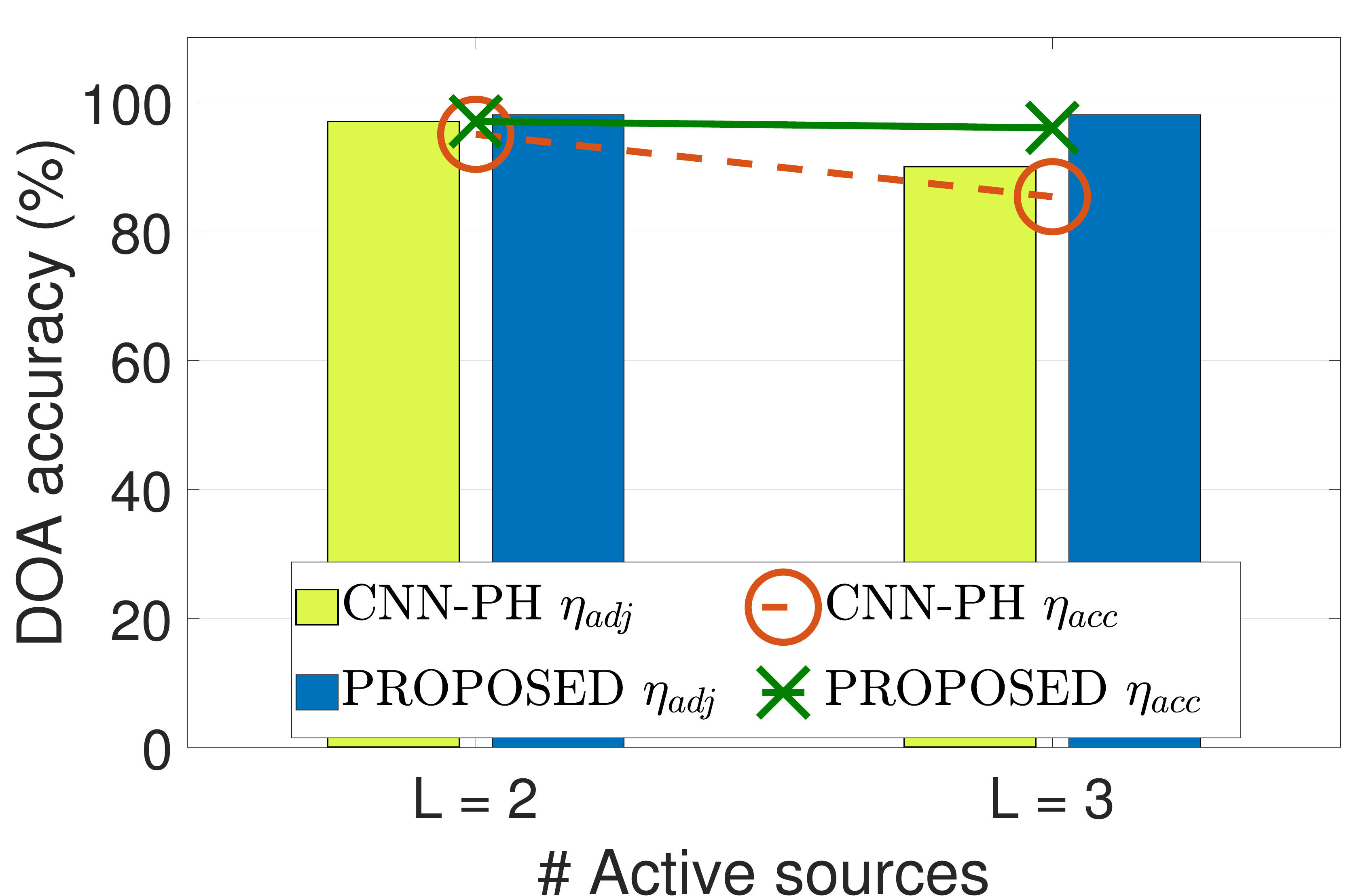}}
\end{minipage}
\caption{Azimuth estimation accuracy with practical recordings with babble noise at $10$dB SNR. The tests were performed on $95^\circ$ elevation plane.}
\label{fig: result-practical}
\end{figure}
Fig. \ref{fig: result-same-plane} shows DOA accuracy of the competing methods under different scenarios. At SNR = $30$dB in Fig. \ref{fig: result-same-plane}, we observe that both the methods perform well for $L = 1$ and $2$ although the proposed method consistently exhibits slightly better performance. For $3$-source combination under stronger reverberation, the proposed method holds the level of adjacent accuracy $\eta_{\text{adj}}$ while \say{CNN-PH} shows performance degradation for both the metrics. For the noisy environments of SNR = $20$dB and $10$dB in Fig. \ref{fig: result-same-plane}, the performance distinctions are more prominent where the proposed algorithm outperforms \say{CNN-PH} in each scenario. The use of modal coherence as learning feature ensures steady performance of the proposed algorithm at low SNR. We can also observe \say{CNN-PH} suffers significant performance issues for $L=3$ due to the fact that we did not train \say{CNN-PH} for all possible $3$-source combinations.\par
For reference, Fig. \ref{fig: histogram-1} plots the TF bin prediction histogram in room S3 for $L=2$ and $3$ along with the true azimuths. The histogram shows a clear peak at each true azimuth location which can be separated using a suitable clustering algorithm.
\subsubsection{\textbf{Performance in a practical room with babble noise}} \label{sec: results-practical}
We conducted the next set of experiments in a big hall with strong reverberation, we named it room P1 in Table \ref{table: test-environments}. The recording was performed with an \textit{Eigenmike} \cite{acoustics2013em32}, however, only first-order harmonics were captured for this task. The source was placed at a $2.8$ m distance from the array in a uniform azimuth grid of $30^\circ$ interval ($J = 12$) on a $95^\circ$ elevation plane. Directional babble noise was added to the recordings at $10$dB SNR from multiple random locations. This time we trained \say{CNN-PH} with all possible angular combinations for both $L=2$ ($132$ angular combinations) and $L=3$ ($1320$ angular combinations) while the proposed method used the same strategy of single-source training for $12$ classes. The comparative performance is shown in Fig. \ref{fig: result-practical} where the proposed algorithm shows a significantly better accuracy than \say{CNN-PH}, especially for $L=3$, despite \say{CNN-PH} being trained for all possible angular combinations in each case. Note that, we found no significant performance improvement for higher harmonic orders. This can be due to low spatial resolution of the training data. The higher order modes can be useful with a denser source distribution,  at high frequencies, or when a regression-based model is used.
\begin{figure}[!t]
\centering
\begin{minipage}[b]{0.99\linewidth}
  \centering
  \centerline{\includegraphics[width=\linewidth]{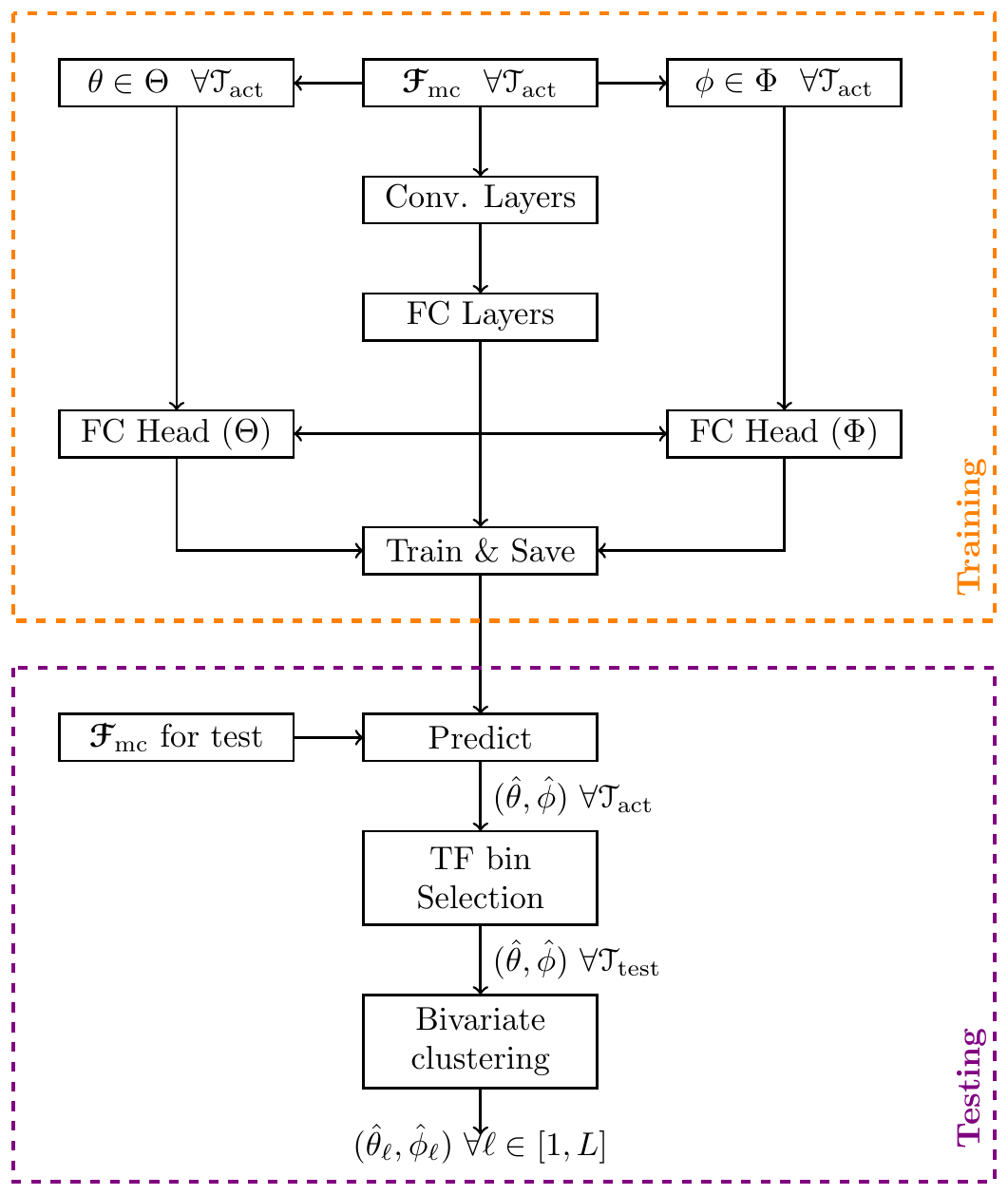}}
\end{minipage}
\caption{A block diagram for joint estimation of azimuth and elevation.}
\label{fig: joint-DOA-block}
\end{figure}
\subsubsection{\textbf{Joint estimation of azimuth and elevation}} \label{sec: result-joint-estimation}
So far, we have shown results only for azimuth estimation based on the proposition that the proposed algorithm can estimate azimuth and elevation simultaneously without interfering with each other. In this section, we are going to validate this proposition by performing a full DOA estimation in room S2. We designed a $3$D uniform spatial grid with $30^\circ$ resolution for azimuths ($J = 12$) and $20^\circ$ resolution for elevations. Furthermore, we considered the elevation range $30^\circ-150^\circ$. That makes a total of $7$ unique elevation classes ($I = 7$) and a total $84$ points on the $3$D DOA grid. The rest of the simulation criteria remain the same as Section \ref{sec: results-sim}.\par
We slightly modified the CNN architecture for this section to accommodate the joint estimation in an efficient manner. As in the previous experiments, we calculated the feature snapshot for each TF bin, but this time we labeled them separately for azimuth and elevation. The CNN architecture remains the same for the most part except at the last layer when we branched out $2$ identical but separated fully connected heads and supplied them with azimuth and elevation labels, respectively. Hence, at the testing stage, the system outputs two separate prediction sets for azimuth and elevation - one from each separated head. Note that, due to the independent estimation strategy, it is important to jointly pack the predicted azimuths and elevations for each TF bin so that the estimated angles from the same source remain together. Subsequently, we clustered the prediction outcomes using the bivariate k-means clustering algorithm. A block diagram for the joint estimation of azimuth and elevation is shown in Fig. \ref{fig: joint-DOA-block}\footnote{Fig. \ref{fig: joint-DOA-block} does not necessarily depict the actual processing flow, rather a visual aid for understanding the task.}. It is worth mentioning that the proposed algorithm can readily be expanded for full source localization through additional training for radius-dependency or corresponding Cartesian coordinates due to its ability of independent estimation of different location parameters.\par
\begin{figure}[!t]
\centering
\begin{minipage}[b]{0.8\linewidth}
  \centering
  \centerline{\includegraphics[width=\linewidth]{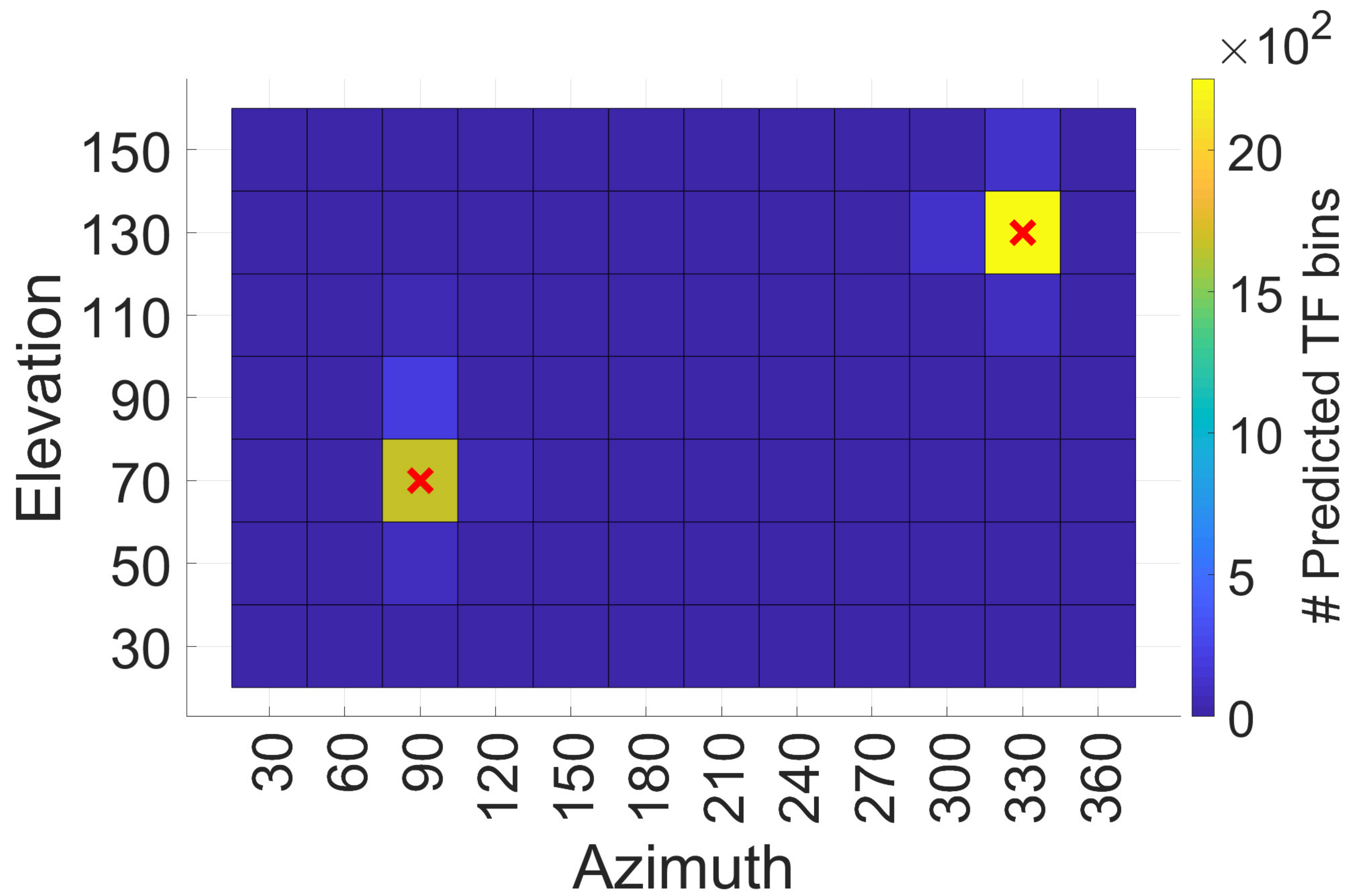}}
  \centerline{{\fontsize{9}{10}\selectfont(a) $L=2$}}\medskip
\end{minipage}
\begin{minipage}[b]{0.8\linewidth}
  \centering
  \centerline{\includegraphics[width=\linewidth]{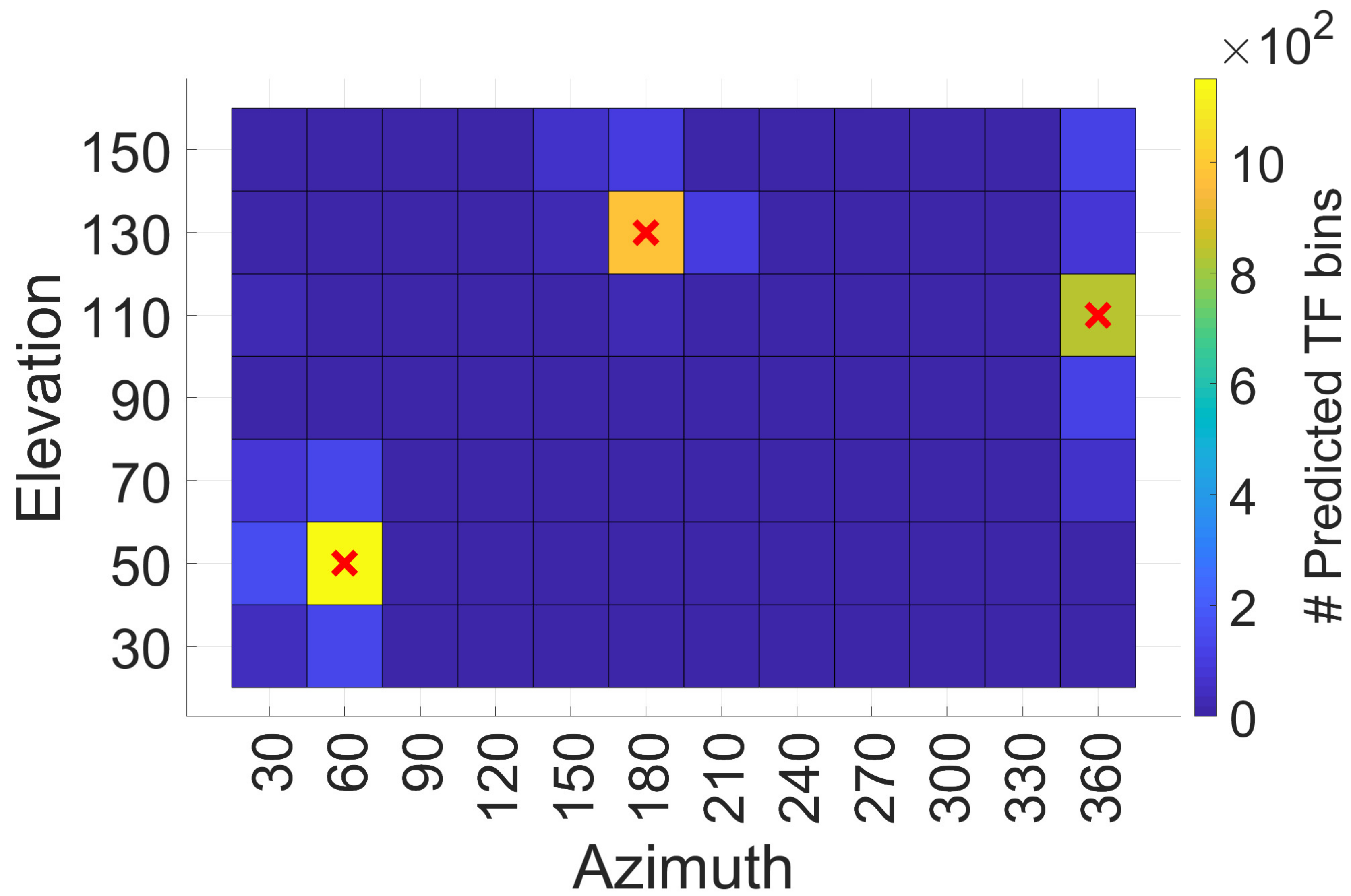}}
  \centerline{{\fontsize{9}{10}\selectfont(b) $L=3$}}\medskip
\end{minipage}
\caption{Color map for joint estimation of azimuth and elevation with the proposed method in room S2 at $30$dB SNR. Red crosses denote the ground truths. The accuracy of a joint estimation over $50$ tests was found to be similar compared to the standalone azimuth estimation.}
\label{fig: result-joint-DOA}
\end{figure}
The outcome of the experiments is shown in Fig. \ref{fig: result-joint-DOA} in terms of colored heat map based on the number of predicted TF bins for each DOA. It is clear from the figure that the proposed method had no difficulties in predicting full DOA in the same manner as with the azimuth predictions. More importantly, the accuracy of the joint DOA estimation was found to be similar to that of a standalone azimuth estimation of Fig. \ref{fig: result-same-plane} (and hence, the accuracy plots are not shown separately for this section). This is expected behavior as the azimuth and elevation estimation processes are independent and should not be affected due to the joint processing.
\subsubsection{\textbf{Azimuth estimation on a different elevation plane}}
\begin{figure}[!t]
\centering
\begin{minipage}[b]{0.80\linewidth}
  \centering
  \centerline{\includegraphics[width=\linewidth]{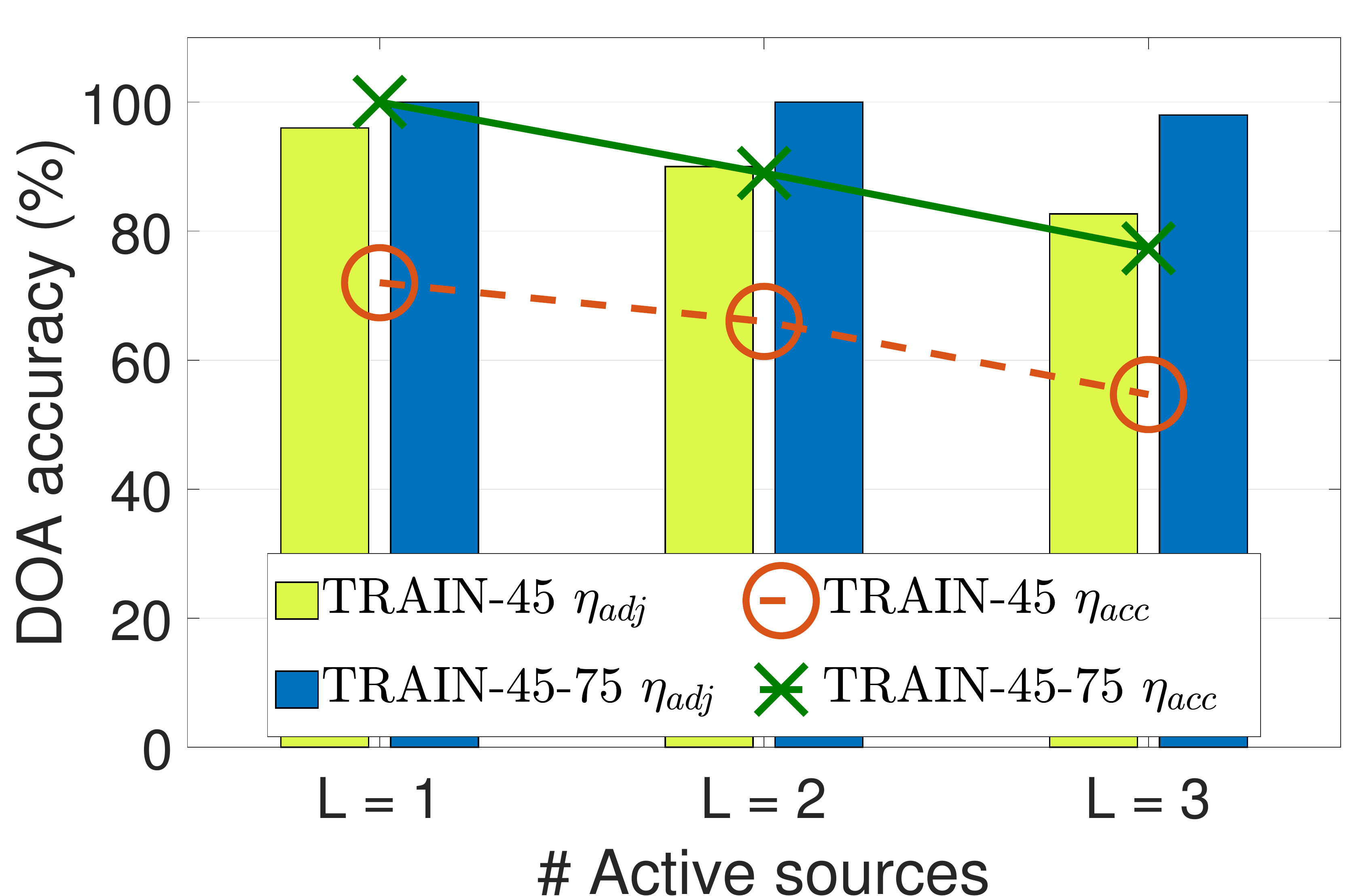}}
\end{minipage}
\caption{Azimuth estimation on $60^\circ$ elevation plane when training was performed on different elevation plane. TRAIN-45 denotes the case when training was performed on $45^\circ$ elevation only whereas for TRAIN-45-75, training was performed with data from $45^\circ$ and $75^\circ$ elevations.}
\label{fig: result-diff-plane}
\end{figure}

In this section, we analyze the performance of the proposed algorithm when training and testing were performed on different elevation planes. For the purpose of this section, we used room S2 at $30$dB SNR. The tests were performed for sources on $60^\circ$ elevation plane while the training data were obtained from a different elevation. In Fig. \ref{fig: result-diff-plane}, we show the estimation accuracy for $2$ distinct cases - when training data were obtained from (case 1) $45^\circ$ elevation plane only and (case 2) $45^\circ$ and $75^\circ$ elevation planes. We observe a clear improvement in case 2 over case 1 due to the fact that when we trained the network on $2$ different elevation planes, the model learned the evolution of feature for change in elevation and predicted azimuths in an arbitrary elevation plane more accurately. As the machine learning algorithms take a data-driven approach, it is possible to further improve the performance by training on additional planes.\par
It is worth noting that the proposed feature snapshot $\fmc$ is mostly comprised of the spherical harmonics where the dependency on $\theta$ and $\phi$ come through independent Legendre and exponential functions, respectively, as shown in \eqref{eq: sh-definition}. Therefore, the impact of elevation change on $\fmc$ comes mainly as a constant scaling factor. For this reason, even for case 1 when training and testing were done in separate individual elevation planes, the model didn't entirely fail, rather gets confused by the reverberation, noise and other non-linear distortions. This is apparent from Fig. \ref{fig: result-diff-plane} where we observe a better accuracy in terms of $\eta_{\text{adj}}$ but a significant difference with $\eta_{\text{acc}}$.\par
\subsubsection{\textbf{Impact of source to microphone distance}}
\begin{figure}[!t]
\centering
\begin{minipage}[b]{0.80\linewidth}
  \centering
  \centerline{\includegraphics[width=\linewidth]{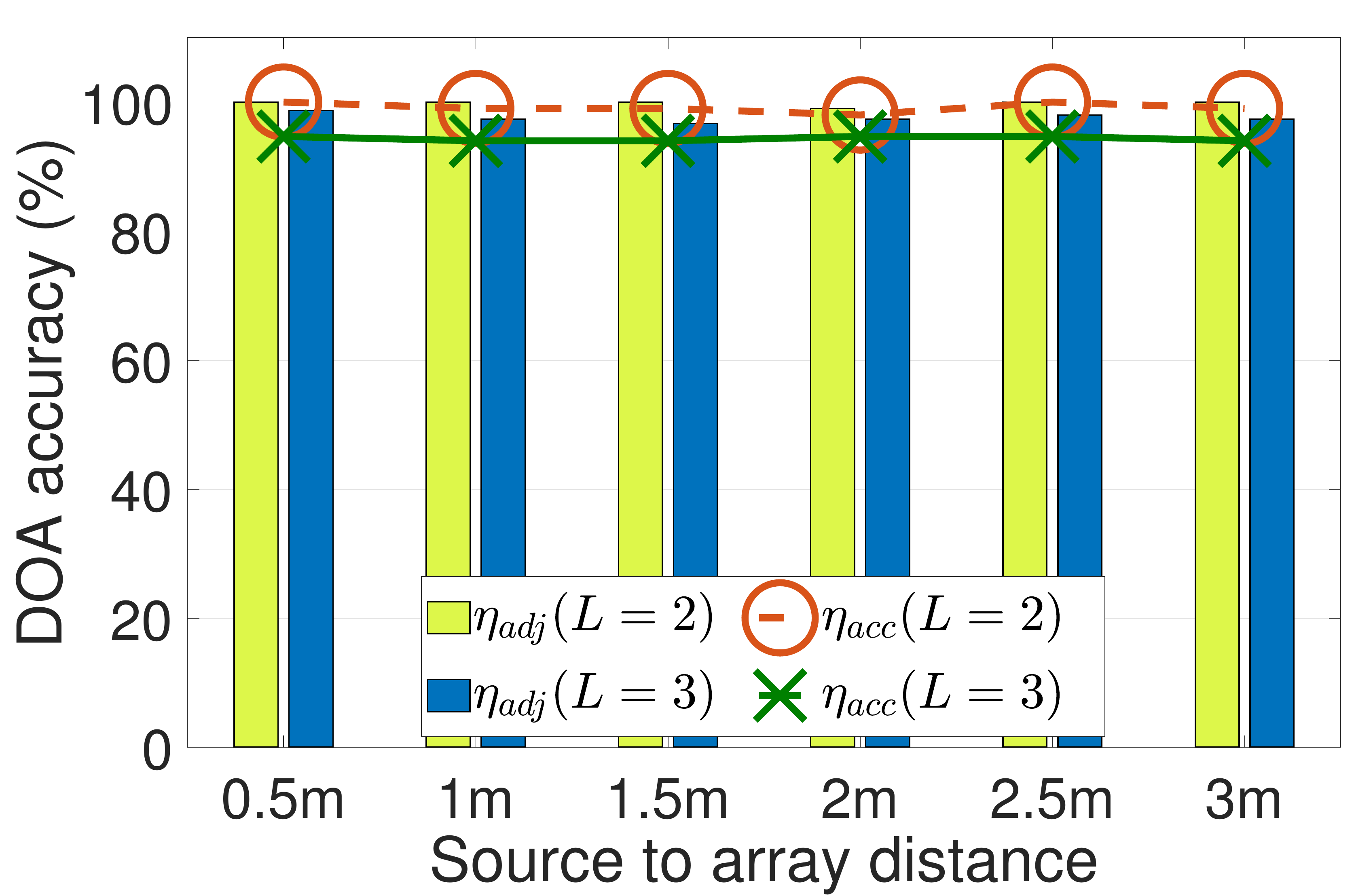}}
\end{minipage}
\caption{Azimuth estimation performance of the proposed algorithm with different source to microphone distances. The training was performed with sources at $1$ m distance from the array center on a $45^\circ$ elevation plane.}
\label{fig: result-diff-s2m}
\end{figure}

We investigate the impact of the varying source to microphone distance on the proposed algorithm. We used the same simulated room S2 with the exception that we increased the dimension of the room to $[8 \times 8 \times 4]$ m for this particular section in order to have a larger range for distance variation. The microphone array position remained at the center of the room, however, we varied the source position between $0.5-3$ m from the microphone array. The training was performed at a fixed distance of $1$ m. The plots in Fig. \ref{fig: result-diff-s2m} suggests that there is no significant change in estimation accuracy for varying source to microphone distances during the test. To understand the behavior, we examine the analytical expression of $\alpha_{nm}$ for the direct path for $\ell^{th}$ source \cite[pp. 31]{colton2012inverse}
\begin{equation} \label{eq: analytical-alpha}
    \alpha_{nm}^{(\ell)}(k, r) = i k h_n(kr_\ell) Y_{nm}^*(\hat{\boldsymbol{x}}_{\ell})
\end{equation}
where $h_n(\cdot)$ is the spherical Hankel function of the first kind. From \eqref{eq: analytical-alpha} it is clear that the radial dependency comes through the Hankel function $h_n(kr_\ell)$ together with the frequency-dependent $k$. As we trained our model for different frequencies, the impact of varying $h_n(kr_\ell)$ on the feature pattern is already captured during the training even with a fixed radius, hence, any radial change does not pose a major threat to the performance of the proposed algorithm.
\subsubsection{\textbf{Number of active sources}}
\begin{figure}[!t]
\centering
\begin{minipage}[b]{0.80\linewidth}
  \centering
  \centerline{\includegraphics[width=\linewidth]{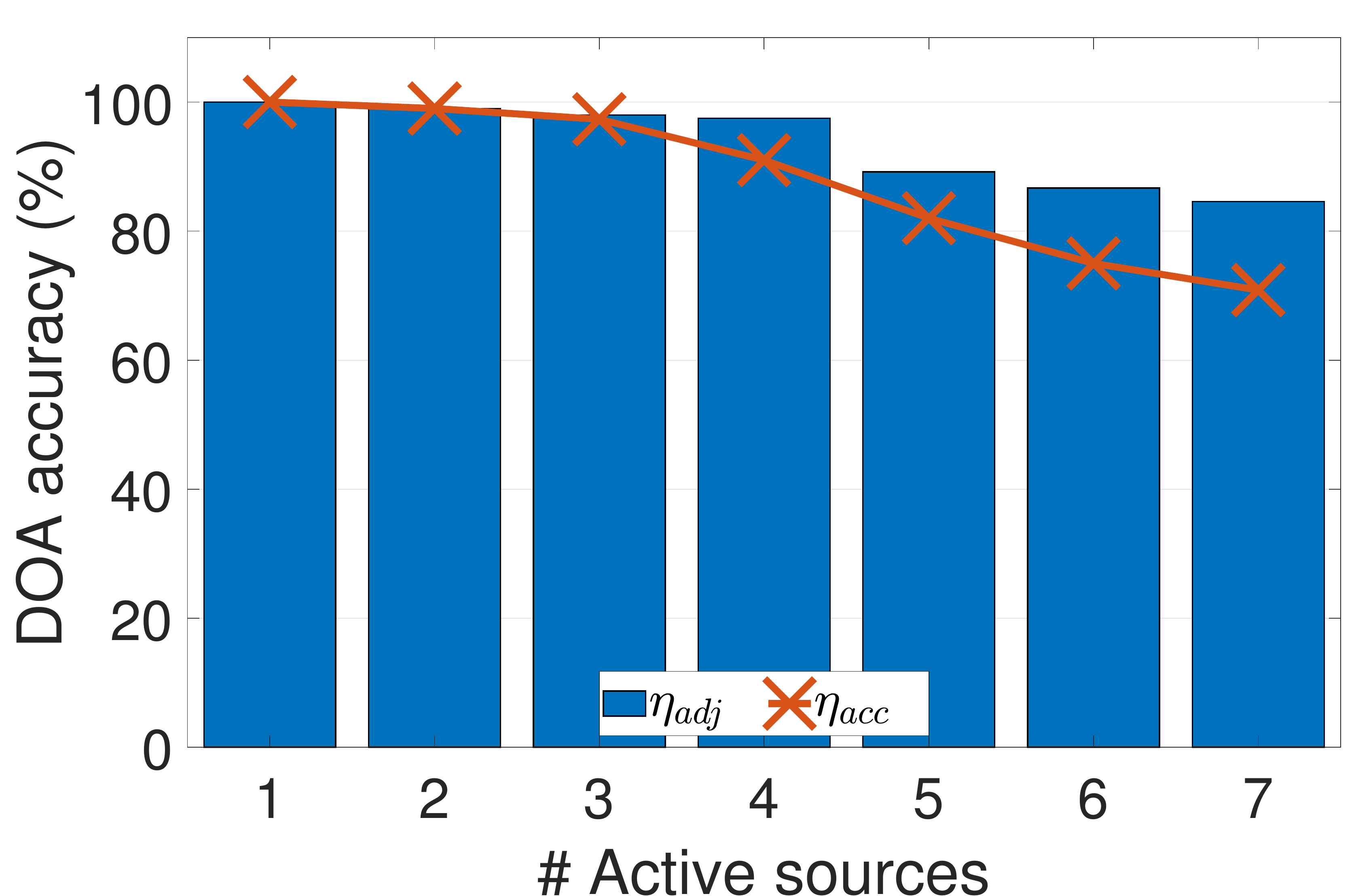}}
\end{minipage}
\caption{Azimuth estimation performance of the proposed algorithm with different number of active sources on a $45^\circ$ elevation plane.}
\label{fig: result-diff-numsources}
\end{figure}

In the last set of experiments, we tried increasing the number of sources on the same $45^\circ$ elevation plane in the acoustic scene of room S2 at $30$dB SNR. As we observe in Fig. \ref{fig: result-diff-numsources}, the accuracy gradually decreases with the increasing number of sources. The performance issue can be contributed by an increased violation of W-disjoint orthogonality with an increased number of sources. However, examining the histograms of random individual tests, we also found many instances when the performance degradation was caused by the failure of the k-means clustering algorithm and the ambiguity in the histogram for nearby sources. It is possible to improve the performance with a careful selection of a more robust clustering algorithm, however, the investigation for a better clustering algorithm is out of the scope of this work. To avoid ambiguity due to nearby sources, we can impose a restriction for maintaining a minimum distance between two sources before applying this algorithm.\par
\section{Conclusions}
In this paper, we proposed a modal coherence based feature to train a convolutional neural network for DOA estimation realized in the spherical harmonic domain. We offered a new perspective by introducing a single-source training scheme for multi-source localization in a reverberant environment. The proposed strategy saves significant time and resources during the training stage as well as allows us to reuse the same trained model during the testing stage irrespective of the number sources in the acoustic mixture. Furthermore, the proposed method is capable of performing parallel azimuth and elevation estimation, hence, allows us to perform full DOA estimation without affecting the estimation accuracy compared to standalone azimuth estimation. Several existing works have already shown that the application of deep learning algorithms in DOA estimation can alleviate the limitations of the parametric approaches such as MUSIC. We further contribute to the cause by proposing a method that performs better than the contemporary CNN-based methods in dynamic acoustic environments, requires significantly less resource and time for training, and predict the DOA based on a single training model for a room irrespective of the number of sources.\par
This work presents multiple future research directions. The proposed technique can be studied with a regression-based prediction model to achieve DOA estimation over a continuous grid. It can also be investigated to develop a generalized model for different reverberant scenarios which is capable of predicting DOA in versatile room environments.
\bibliography{abdfahim}
\bibliographystyle{IEEEtran}
\begin{IEEEbiography}[{\includegraphics[width=1in,height=1.25in,clip,keepaspectratio]{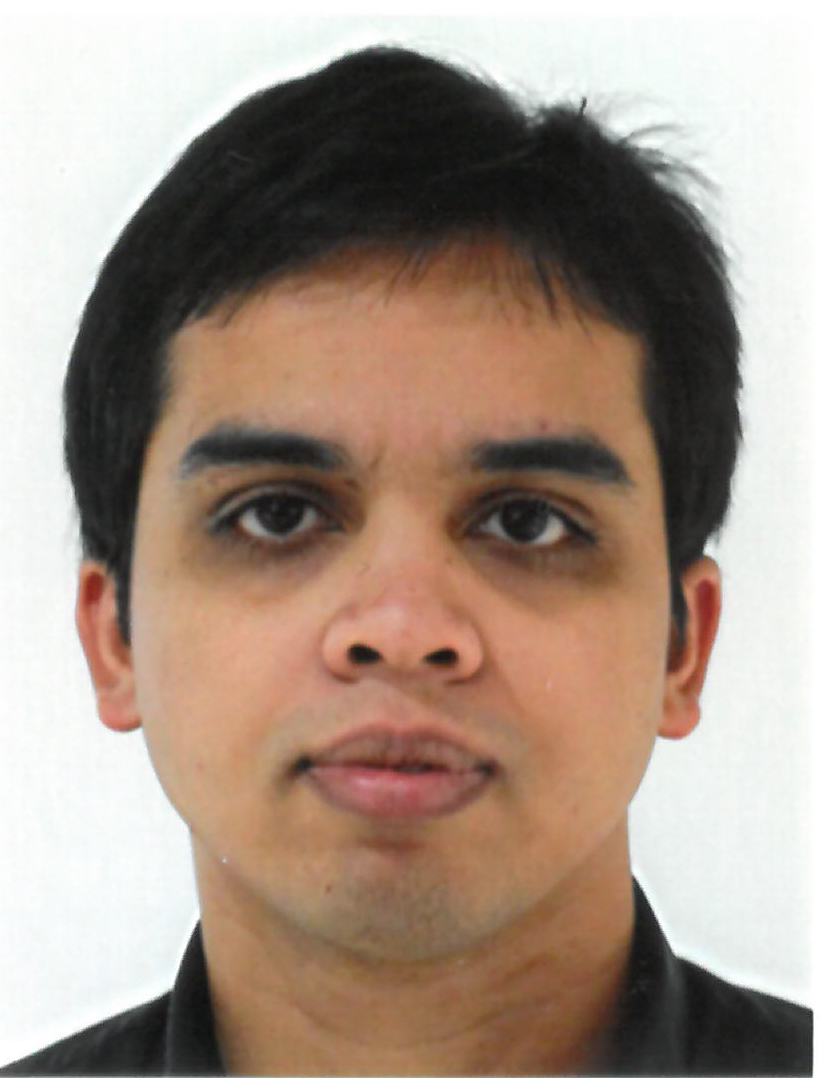}}]{Abdullah Fahim}
received the B.Sc.(Hons.) degree in electrical and electronic engineering from Bangladesh University of Engineering and Technology, Dhaka, Bangladesh, in 2007. From 2007-2015, he was involved in different projects with Ericsson and Nokia SN. He is currently pursuing his Ph.D. degree in spatial audio signal processing from the Australian National University, Canberra, Australia. In 2018, he worked with the spatial audio team of Apple Inc. in California for a $6$-months' internship program. His research interests include spatial audio processing techniques, virtual audio, and soundfield separation and enhancement.
\end{IEEEbiography}
\begin{IEEEbiography}[{\includegraphics[width=1in,height=1.25in,clip,keepaspectratio]{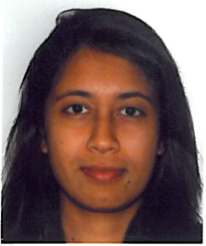}}]{Prasanga N. Samarasinghe}
received  her Ph.D. degree from the Australian National University (ANU), Australia in 2014 and her B.E. degree (with Hons.) in electronic and electrical engineering from the University of Peradeniya, Sri Lanka in 2009. In 2019 she received a prestigious Fulbright Future Scholarship to visit the University of Maryland, USA. She is currently a Senior Lecturer at ANU, and her research interests include spatial sound recording, reproduction and analysis, room acoustics, spatial noise cancellation and virtual audio.
\end{IEEEbiography}
\begin{IEEEbiography}[{\includegraphics[width=1in,height=1.25in,clip,keepaspectratio]{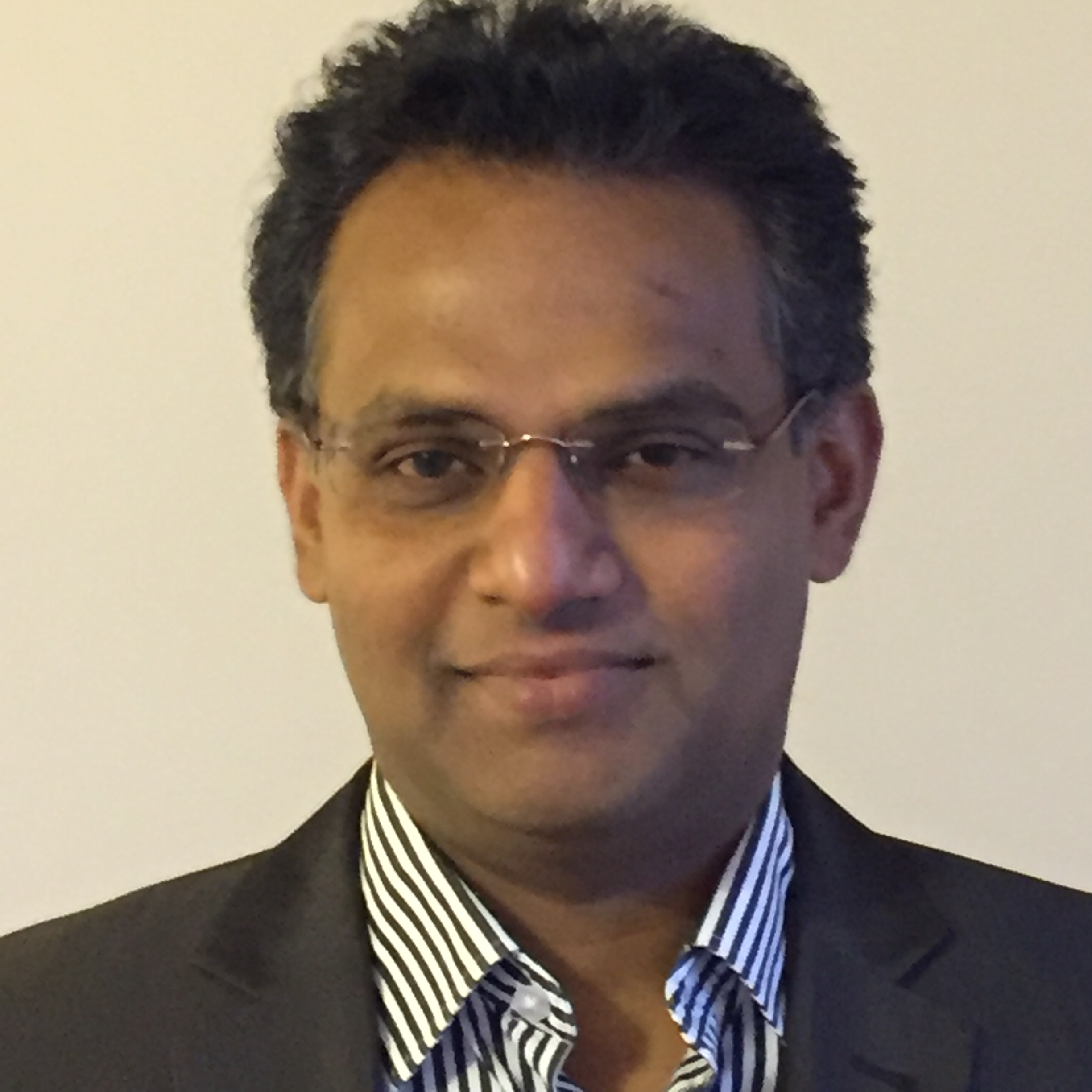}}]{Thushara D. Abhayapala}
is a Professor of Signal Processing at the Australian National University (ANU), Canberra. He received his B.E. degree in engineering in 1994 and his Ph.D. degree in telecommunications engineering in 1999, both from the ANU. He held a number of leadership positions including Deputy Dean of the College of Engineering and Computer Science (2015-19), Head of the Research School of Engineering at ANU (2010-14) and the leader of the Wireless Signal Processing Program at the National ICT Australia (NICTA) from 2005-07. His research interests are in the areas of spatial audio and acoustic signal processing, and multichannel signal processing. Among many contributions, he is one of the first researchers to use spherical harmonic based eigen-decomposition in microphone arrays and to propose the concept of spherical microphone arrays; novel contributions on the multi-zone soundfield reproduction problem; was one of the first to show the fundamental limits of spatial soundfield reproduction using arrays of loudspeakers and spherical harmonics. He worked in industry for two years, before his doctoral study and has active collaboration with a number of companies. He has supervised 36 PhD students and co-authored more than 280 peer-reviewed papers. He was an associate editor of IEEE/ACM Transactions on Audio, Speech, and Language Processing and was a member of the Audio and Acoustic Signal Processing Technical Committee (20112016) of the IEEE Signal Processing Society. He is a fellow of Engineers Australia (IEAust).
\end{IEEEbiography}
\end{document}